%% file: MAIN.tex
\documentclass[a4paper]{paper_cw}

\usepackage{helvet}
\usepackage{amsmath}
\usepackage{amssymb}
\usepackage{mathrsfs}
\usepackage{graphics}
\usepackage[numbers]{natbib}
\usepackage{color}
\usepackage{fancyhdr} 
\usepackage{colortbl} 
\usepackage[dvips]{graphicx} 
\usepackage{marvosym}
\usepackage{float}

\usepackage{array} 
\usepackage{framed}
\usepackage{subcaption}

\usepackage{pgfplots}
\usepackage{tcolorbox}
\usepackage{multirow}
\usepackage{bm}
\usepackage{url}
\usepackage{relsize}
\usepackage{ulem}

\usepackage{listings}
\definecolor{deepblue}{rgb}{0,0,0.5}
\definecolor{deepred}{rgb}{0.6,0,0}
\definecolor{deepgreen}{rgb}{0,0.5,0}

\usepackage[colorlinks=true, allcolors=blue]{hyperref}

\begin{document}
\include{definitions}


\title{Physics-Informed Surrogates for Temperature Prediction of Multi-Tracks in Laser Powder Bed Fusion}

\author{H. Safari, $^{1}$ (\Letter), H. Wessels$^{1}$}

\institute{(1) Institute of Applied Mechanics, Technische Universtität Braunschweig, Pockelsstraße 3, 38106 Braunschweig
\email{hesameddin.safari@tu-braunschweig.de}\\}

\maketitle
\thispagestyle{empty}

\abstract{Modeling plays a critical role in additive manufacturing (AM), enabling a deeper understanding of underlying processes. Parametric solutions for such models are of great importance, enabling the optimization of production processes and considerable cost reductions. However, the complexity of the problem and diversity of spatio-temporal scales involved in the process pose significant challenges for traditional numerical methods. Surrogate models offer a powerful alternative by accelerating simulations and facilitating real-time monitoring and control. The present study presents an operator learning approach that relies on the deep operator network (DeepONet) and physics-informed neural networks (PINN) to predict the three-dimensional temperature distribution during melting and consolidation in laser powder bed fusion (LPBF). Parametric solutions for both single-track and multi-track scenarios with respect to tool path are obtained. To address the challenges in obtaining parametric solutions for multi-track scenarios using DeepONet architecture, a sequential PINN approach is proposed to efficiently manage the increased training complexity inherent in those scenarios. The accuracy and consistency of the model are verified against finite-difference computations. The developed surrogate allows us to efficiently analyze the effect of scanning paths and laser parameters on the thermal history.
}

\keywords{physics-informed neural networks, operator learning, additive manufacturing, powder bed fusion}

\section{Introduction}\label{sec:Introduction}
Additive manufacturing (AM) provides great opportunities to efficiently design and fabricate individualized components. It also offers unparalleled flexibilities for fabricating metamaterials that are impossible to create with conventional processes. However, the complexities of AM processes have posed great challenges in maintaining consistent quality, necessitating careful optimization of process parameters. For instance, in Laser Powder Bed Fusion (LPBF), various process parameters play a crucial role. These include laser characteristics such as power, spot size, and scanning speed, as well as scan strategies like hatch spacing and scan patterns, to name just a few. Each parameter influences the quality, precision, and mechanical properties of the final product.

Numerical simulations based on governing partial differential equations (PDEs) provide us with valuable information and insights about the underlying physics which gives us the possibility to perform optimization studies.  AM is often characterized by multi-scale, complex, and rapid physical phenomena that can span more than 3–4 orders of magnitude in both spatial resolution and temporal scales~\cite{markl2016multiscale}. This vast range presents significant challenges in terms of computational resources and algorithmic complications with established numerical schemes, such as finite-element method (FEM)~\cite{schoinochoritis2017simulation, sarkar2024advances}, finite-difference method (FDM)~\cite{foteinopoulos2018thermal}  and finite-volume method (FVM)~\cite{li2023efficient}. Thus, parametric solutions and parametric studies in terms of process parameters that are needed for conducting such optimizations are not feasible using classical numerical algorithms for real applications. Surrogate modeling can play an important role in bridging the gap between the need for computationally efficient solutions and the complexity of numerical simulations. Surrogate models allow for the rapid exploration of a wide parameter space, making optimization studies feasible even for computationally intensive AM simulations. In addition, they facilitate real-time predictions and process adjustments by significantly reducing computational requirements.

The reduced-order modeling approach is one of the most widely used methods in AM for development of surrogate models, especially for computation of the temperature history. Zheng \textit{et al.}~\cite{zheng2017modeling} have presented a method for modeling and controlling the cooling rate in laser additive manufacturing. The partial differential equation governing the process is transformed and approximated using the proper orthogonal decomposition (POD) method. The study also explores cooling rate control strategies, including feed-forward control based on steady-state solutions and feedback control to maintain desired thermal profiles. POD-based models rely on the linear superposition of modes, which may struggle to accurately capture complex, nonlinear dynamics, particularly in the context of nonlinear, time-dependent, parameterized PDEs~\cite{fresca2021comprehensive, agarwal2024parameter}. As a remedy, proper generalized decomposition (PGD) model has been developed by Favoretto \textit{et al.}~\cite{favoretto2019reduced} to account for highly transient temperature evolutions in LPBF. They reported a simulation speed-up of at least one order of magnitude relative to standard techniques. However, significant challenges arise when applying PGD to transient thermal problems involving a moving heat source. The performance of PGD models is highly sensitive to factors such as the choice of discretization method, the size of the heat source, and material parameters~\cite{strobl2024pgd}. Addressing these challenges often requires problem-specific treatments to ensure accuracy and stability of the model.


On the other hand, machine learning (ML)-based methods offer significant advantages for presenting surrogates for AM. Their ability to model complex, nonlinear relationships, enables accurate representation of complicated process dynamics. By integrating multi-fidelity data from various sources, these models enhance prediction accuracy and reliability~\cite{demo2023deeponet}. Once trained, ML models deliver real-time predictions, supporting timely monitoring and control of manufacturing processes. For instance, Yaseen \textit{et al.}~\cite{yaseen2023fast} have developed fast and accurate reduced order models for AM processes using operator learning techniques, such as the Fourier Neural Operator (FNO~\cite{li2020fourier}) and Deep Operator Network (DeepONet~\cite{lu2021learning}). These methods effectively predict time-dependent responses in AM models, enhancing process optimization. Chen \textit{et al.}~\cite{chen2024capturing} proposed a data-driven model using a Fourier neural operator to capture local temperature evolution in wire-based directed energy deposition (DED) AM. Their results showed that the model accurately predicts temperature changes regardless of geometry. However, the main limitation of the presented model is that it is only applicable to a single toolpath. In~\cite{kim2022tool}, a convolutional neural network (CNN) has been trained based on the numerical simulations in order to predict the optimal laser paths during selective laser sintering (SLS) process. Supervised data-driven ML models require large, high-quality datasets for training. Obtaining such data can require significant resources for AM processes. Moreover, the accuracy and generalizability of the models heavily depend on the quality and diversity of the available data.

Physics-informed neural networks (PINNs)~\cite{raissi2019physics} represent an innovative approach which addresses the limitations of purely data-driven deep learning frameworks. By incorporating underlying physics (in the form of well-known PDEs or other physical constraints) in the loss function, PINNs ensure that the model respects the established governing laws. This integration not only enhances the interpretability of the network but also significantly improves its generalization capabilities, even in the absence of extensive simulation/experimental data. Nevertheless, data can support the training process, see~\cite{anton2024deterministic}. The paradigm of PINNs has recently captured the attention of the AM community as a promising tool for developing parametric surrogate models. The efforts in this direction has been initiated by the pioneering work of Zhu \textit{et al.}~\cite{zhu2021machine}, in which they employed the Navier-Stokes equations together with the energy equation into a PINN to compute the temperature and meltpool dynamics in LPBF. They demonstrated that the PINN can accurately predict the meltpool dimensions and the temperature distribution using only a limited amount of labeled training data. Nevertheless, they only considered a forward non-parametric solution for the problem in question. Xie \textit{et al.}~\cite{xie20223d} have solved the transient heat conduction equation with a PINN to compute the three-dimensional temperature fields in single-layer and multi-layer DED processes. Similarly, Li \textit{et al.}~\cite{li2023physics} presented a PINN framework for temperature prediction in laser metal deposition processes without labeled data. Both deposition and cooling stages have been considered in their study. They employed transfer learning across scenarios with varying manufacturing parameters to enable faster predictions while maintaining accuracy. 

Liao \textit{et al.}~\cite{liao2023hybrid} have developed a hybrid thermal modeling approach for a DED process that integrates physics-based principles with data-driven techniques using PINNs. By combining partially observed temperature data from infrared camera measurements with fundamental physics laws, the method enables the prediction of full-field temperature histories and the identification of unknown material properties (including the heat capacity and the material thermal conductivity) as well as process parameters such as laser absorptivity. In another study, Hosseini \textit{et al.}~\cite{hosseini2023single} investigated the applicability of parametric PINNs in order to obtain the parametric solutions of LPBF process over a wide range of material properties and process parameters. They reported a Mean Absolute Error of below 5\% for the temperature and the meltpool sizes relative to FEM solutions for all the considered benchmark tests. In later work, the authors integrated a parametric PINN with a cellular automata (CA) microstructure model to calibrate a thermo-microstructural model based on single-track LPBF experiments~\cite{tang2024calibration}. The trained PINN provided parametric solutions in terms of process conditions and heat source model parameters. To minimize the discrepancies between simulation and experiment, they conducted an inverse analysis to optimize the thermal model's unknown parameters. A recent review on different approaches for the calibration of numerical models (including PINN) from full-field data can be found in \cite{romer2024reduced}.

All the aforementioned PINN frameworks have been limited to single-track AM processes. To the best of our knowledge, no foundational study has explored the application of PINNs for multi-track AM processes. This is particularly important since tool path selection directly impacts build quality, mechanical properties, and thermal management~\cite{kim2022tool,flores2019toolpath}. Efficient paths also reduce build time, material waste and improve overall process efficiency~\cite{xiong2019process}. Hence, in this research, we explore the potential of developing PINN surrogate models and physics-informed neural operators for multi-track laser AM applications. We pay special attention to neural operators (more specifically, DeepONet), due to its robust theoretical foundation for approximating operators that map between function spaces. This makes DeepONet a promising candidate to create surrogate solvers for parametric PDEs. Within this framework, each laser path can be considered as an input function to the PDE of the heat equation. We compare the physics-informed DeepONet (PI-DeepONet) with PINN in terms of the their performance and accuracy for multi-track scenarios. Moreover, we propose a sequential-PINN framework to efficiently account for different laser paths.

The rest of this paper is structured as follows. In Section~\ref{sec:methodology}, we first describe the numerical implementation of the heat equation using a finite-difference scheme and then present the physics-informed deep learning frameworks, including the formulation of the PI-DeepONet and a sequential PINN strategy for multi-track scenarios. In Section~\ref{sec:result}, we report and discuss the results obtained for both single-track and multi-track cases, comparing the surrogate models with reference finite-difference solutions in terms of temperature distributions and meltpool dimensions. Finally, Section~\ref{sec:conclusion} concludes with a summary of the key findings, along with an outlook on future work.

\section{Methodology} \label{sec:methodology}

The heat equation that describes the distribution of temperature in a given region over time is derived from the conservation of energy. Neglecting the convective and heat dissipation loss terms, the partial differential equation for the evolution of temperature in time-space domain can be written as, 
\begin{equation}\label{eq:heat}
    \rho c_p \partial_t T + \nabla \cdot  \tenq = q_v,
\end{equation}
where $\rho$ denotes the density, $c_p$ the heat capacity, $T$ the temperature, $\tenq$ the heat flux and $q_v$ a volumetric heating term. For the heat flux, we consider Fourier type heat conduction with $\tenq = - \kappa(T(\vecx)) \nabla T$, where $\kappa$ is temperature-dependent thermal conductivity. Thus, \eqref{eq:heat} can be written as
\begin{equation}\label{eq:temperature}
    \rho c_p \partial_t T = \kappa \nabla^2 T + \nabla \kappa \cdot \nabla T + q_v.
\end{equation} 
The heat equation \eqref{eq:temperature} is subject to Dirichlet and Neumann boundary conditions as follows
\begin{equation}
    \begin{aligned}
        T &= \bar{T} \, &\text{on} \, \Gamma^{\theta}_D \\
        \vecq \cdot \vecn &= \bar{q} \, &\text{on} \, \Gamma^{q}_N.
    \end{aligned}
\end{equation}
The implementation of the laser heating can be either achieved via a volumetric heat source or application of a proper heat flux as the boundary condition on the top surface. In this study, the volumetric heat source is applied using the moving Gaussian point heat source. Assuming a hemispherical Gaussian distribution of power density for a moving laser in $xy$ plane, $q_v$ is given by~\cite{hosseini2023single},
\begin{equation}
    \label{eqn:qlaser}
    \begin{aligned}
        q_v(x,y,z,t) = \eta \frac{6 \sqrt{3} P}{r^3 \pi \sqrt{\pi}}
        \text{exp} \left( {-3{\frac{(x - x_l)^2 + (y - y_l)^2 }{r^2}}} \right)
        \text{exp} \left( {-3{\frac{z^2 }{c^2}}} \right),
    \end{aligned}
\end{equation}
where $P$ represents the laser power, $\eta$ the laser absorption coefficient, $x_l$ and $y_l$ the location of the laser beam in $xy$ plane, $r$ the laser radius and $c$ the laser penetration depth.

\subsection{Solving Heat Equation using Finite-Difference Method (FDM)} \label{subsec:fdm}

Discretizing \eqref{eq:temperature} using explicit finite difference with a forward Euler scheme in time, the temperature at each time step can be computed by,
\begin{equation}
    \label{eqn:fd}
    \begin{aligned}
        T(t+\delta t,x,y,z) &= T(t,x,y,z) + \frac{\delta t}{\rho c_p} \left[
        \kappa(x,y,z)\left( D_x + D_y + D_z\right) + K_x + K_y + K_z + q_v
        \right] \\
    \end{aligned}
\end{equation}
where
\begin{equation}
    \begin{aligned}
        D_x &= \frac{\partial^2 T}{\partial x^2} = \frac{T(t,x + \delta x,y,z) - 2T(x,y,z) + T(t,x - \delta x,y,z)}{\delta x^2} \\
        D_y &= \frac{\partial^2 T}{\partial y^2} = \frac{T(t,x,y+\delta y,z) - 2T(x,y,z) + T(t,x,y - \delta y,z)}{\delta y^2}\\
        D_z &= \frac{\partial^2 T}{\partial z^2} = \frac{T(t,x,y,z+\delta z) - 2T(x,y,z) + T(t,x,y,z-\delta z)}{\delta z^2} \\
        K_x = \frac{\partial \kappa}{\partial x}\frac{\partial T}{\partial x} &= \frac{\kappa(x + \delta x,y,z) - \kappa(x - \delta x,y,z)}{2\delta x} \cdot \frac{T(t,x + \delta x,y,z) - T(t,x - \delta x,y,z)}{2\delta x}   \\
        K_y = \frac{\partial \kappa}{\partial y}\frac{\partial T}{\partial y} &= \frac{\kappa(x,y + \delta y,z) - \kappa(x,y - \delta y,z)}{2\delta y} \cdot \frac{T(t,x,y +\delta y,z) - T(t,x,y - \delta y,z)}{2\delta y}  \\
        K_z = \frac{\partial \kappa}{\partial z}\frac{\partial T}{\partial z} &= \frac{\kappa(x,y,z+\delta z) - \kappa(x,y,z-\delta z)}{2\delta z} \cdot \frac{T(t,x,y,z+\delta  z) - T(t,x,y,z - \delta z)}{2\delta z} .
    \end{aligned}
\end{equation}
In this manuscript, we use solutions obtained with finite differences as a baseline to benchmark the physics-informed surrogates under consideration.

\subsection{Physics-informed neural network architectures as heat equation solvers}\label{subsec:pi-deeponet}

\noindent \textbf{Standard PINN:} The core idea behind the PINN is to construct a deep neural network (DNN) that serves as a surrogate model (ansatz function) to the solution of the PDE. By incorporating appropriate activation functions, the output of the network is differentiable with respect to the inputs (here the spatiotemporal coordinates)~\cite{raissi2019physics, mao2020physics}. This enables the automatic differentiation of the required derivatives in order to introduce the underlying physics, namely the heat equation~\eqref{eq:temperature} as well as the corresponding boundary conditions. Implementing the latter in terms of a loss function, one can enforce the network to satisfy the governing PDEs by minimization of the total loss function. This enables us to estimate the PDE solution without any prior knowledge about the temperature distribution in an unsupervised manner. Hence, the learned solution remains consistent with the physics and mathematical models. This especially is of great interest, as obtaining data from Finite-Element (FE) simulations of AM applications for data-driven surrogate models could be cumbersome in terms of computational resources~\cite{hosseini2023single}. 
In this context, collocation points across the domain are required to evaluate and enforce the governing physical laws. The low-discrepancy Sobol sampling method~\cite{sobol1967distribution} is employed for sampling the collocation points in time-space domain. Each collocation point can be viewed as a four-component vector in the form of $(t_i, x_i, y_i, z_i)$ or $(t_i, \mathbf{\hat{x}}_i)$. As suggested by Hosseini~\cite{hosseini2023single}, we use an adaptive clustering strategy in the vicinity of the heat source to ensure capturing sharp temperature gradients. The initial and boundary conditions and the PDE are enforced with the loss functions defined as follows,


\begin{align}
    \mathcal{L}_\text{IC} &= \frac{1}{N_{IC}} \sum_{i=1}^{N_{IC}} \left| \mathcal{T}(t=t_0, \mathbf{\hat{x}}_i) - T_{0}(\mathbf{\hat{x}}_i) \right|^2, \label{eq:loss_ic} \\
    \mathcal{L}_{ \text{BC,Dirichlet}} &= \frac{1}{N_{BC}} \sum_{i=1}^{N_{BC}} \left| \mathcal{T}(t_i, \mathbf{\hat{x}}_i) - T_{BC}(t_i, \mathbf{\hat{x}}_i) \right|^2, \label{eq:loss_bc_d}  \\
    \mathcal{L}_{\text{BC,Neumann}} &= \frac{1}{N_{BC}} \sum_{i=1}^{N_{BC}} \left| \frac{\partial}{\partial n} \mathcal{T}(t_i, \mathbf{\hat{x}}_i) - \frac{\partial}{\partial n} T_{BC}(t, \mathbf{\hat{x}}_i)\right|^2, n \in \{x, y, z\}, \label{eq:loss_bc_n} \\
    \mathcal{L}_{\text{PDE}} &=  \frac{1}{N_{PDE}} \sum_{i=1}^{N_{PDE}} \left| \mathcal{R}_{PDE} (t_i, \mathbf{\hat{x}}_i) \right|^2, \label{eq:loss_pde}
\end{align}
where $N_\bullet$ is the total number of time-space data points in each dataset.
$\mathcal{T}(t, \mathbf{\hat{x}})$ represents the output of the network, $T_0$ the value of the initial condition, $T_{BC}$ the assigned temperature on the related boundary and $\mathcal{R}_\text{PDE}$ the residuals of the governed PDE. To satisfy the heat equation on the collocation points inside the computational domain, the residuals of the PDE for the $i^{th}$ sample in the collocation points dataset is calculated as,
\begin{align} \label{pde_res}
    \mathcal{R}_\text{PDE}(t_i, \mathbf{\hat{x}}_i) &=  \rho c_p \frac{\partial \mathcal{T}(t_i, \mathbf{\hat{x}}_i)}{\partial t} - \frac{\partial}{\partial x} \left( \kappa \frac{\partial \mathcal{T}(t_i, \mathbf{\hat{x}}_i)}{\partial x} \right) - \frac{\partial}{\partial y} \left( \kappa \frac{\partial\mathcal{T}(t_i, \mathbf{\hat{x}}_i) }{\partial y} \right) \\
    & - \frac{\partial}{\partial z} \left( \kappa \frac{\partial \mathcal{T}(t_i, \mathbf{\hat{x}}_i)}{\partial z} \right)  - q_v(t_i, \mathbf{\hat{x}}_i).
\end{align}

Summing up the above-mentioned loss terms, the final loss function is given by,
\begin{equation}
    \label{loss_total}
    \mathcal{L}_\text{Total} =  \lambda_\text{IC}\mathcal{L}_\text{IC} +  \lambda_\text{BC}\sum_{i=1}^{M} \mathcal{L}_{\text{BC},i} +  \lambda_\text{PDE}\mathcal{L}_\text{PDE},
\end{equation}
with $M$ being the total number of boundary conditions. The regularization of total loss term is achieved by means of different weighting factors ($\lambda$) for each individual loss term. The values of the weighting factors are computed as follows,
\begin{equation}\label{eq:lambda}
\begin{aligned}
    \lambda_\text{IC} = \frac{1}{\mathcal{L}_{0,\text{IC}}} \\
    \lambda_\text{BC} = \frac{1}{\sum_{i=1}^{M} \mathcal{L}_{0,\text{BC},i}} \\
    \lambda_\text{PDE} = \frac{1}{\mathcal{R}_{0,\text{PDE}}}
\end{aligned}
\end{equation}
where the subscript '0' denotes the initial loss terms at the beginning of the training process at $t=0$.
\\

\noindent \textbf{Physics-informed En-DeepONet (PI-EnDeepONet):} Non-parametric forward PINNs as introduced above utilize time-space coordinates as input to the network. This architecture can also be extended to explore parametric solutions by incorporating additional dimensions into the input tensor, enabling PINNs to learn solutions across time-space-parameters domain~\cite{hosseini2023single, cao2024solving, panahi2025modeling}. For instance, for a fixed tool path in the laser AM process one can incorporate the process parameters (laser power, scan speed) as well as the material parameters (thermal conductivity, heat capacity etc.) into the network input along side the time-space data points to infer a parametric solution~\cite{hosseini2023single}. These parameters can be more or less seen as coefficients in the heat equation. However, if we aim to obtain parametric solutions for multiple laser paths or scan patterns, these constitute input functions.
Therefore, we require networks capable of mapping input functions to output functions. Different neural operators have been proposed recently to learn mapping across function spaces including DeepONet~\cite{lu2021learning}, Fourier Neural Operator (FNO)~\cite{li2020fourier} and  wavelet neural operator (WNO)~\cite{tripura2023wavelet}.

\begin{figure}[t]
    \centering
    \includegraphics[width=0.50\linewidth]{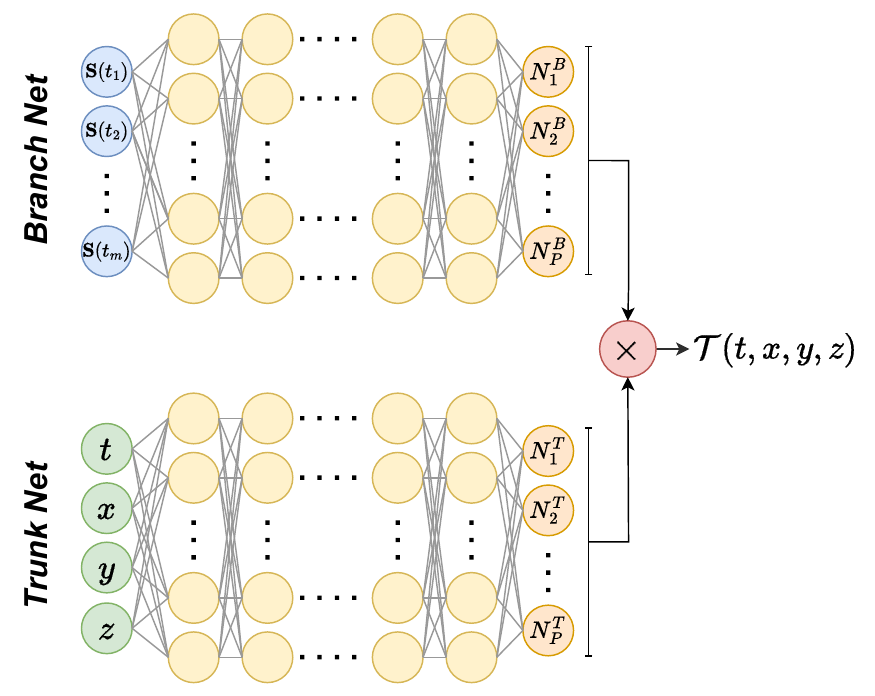}
    \caption{Schematic representation of the original DeepONet.}
    \label{fig:deeponet}
\end{figure}

Here, the DeepONet architecture is selected to obtain a spatiotemporal surrogate for predicting temperature in AM processes, including multi-track scenarios. Fig.~\ref{fig:deeponet} schematically illustrates the original DeepONet architecture \cite{lu2021learning} for the problem in question. The trunk net takes spatiotemporal coordinates are denoted by $(\mathbf{\hat{x}},t)$, where $\mathbf{\hat{x}} = (x, y, z)$ denotes the spatial coordinates and $t$ the time. The tool path (or laser path) for each scenario serves as the input function, $\mathbf{x}_l$, to the branch net. This input function should be represented discretely at fixed locations for all samples. Therefore, the coordinates of the path which the heat source travels during the process are discretized at fixed points in the time domain. The input function $\mathbf{x}_l$ is defined as,

\begin{equation}\label{eq:input_function}
    \begin{aligned}
    \mathbf{x}_l(t) &= \int_t \mathbf{v}_l(t) \, \text{d}t,
    \end{aligned}
\end{equation}
with $\mathbf{v}_l$ being the laser velocity vector on the $xy$-plane and $t$ the process time. Considering $n$ different scenarios for the tool path and $m$ equally-spaced points (or time steps) within the process time, the input function of $i^{th}$ scenario is represented as follows,
\begin{equation}
    \label{eq:s_i}
    \begin{aligned}
    \Big\{ \mathbf{x}_l^{(i)}(t_1), \mathbf{x}_l^{(i)}(t_2), \ldots, & \mathbf{x}_l^{(i)}(t_m) \Big\}.
    \end{aligned}
\end{equation}
According to Fig.~\ref{fig:deeponet}, the output of the network is computed by,
\begin{equation}\label{eq:output}
    \mathcal{T}(\mathbf{x}_l)(\mathbf{\hat{x}},t) \approx \sum_{k=1}^{p} \underbrace{N^B_k\big(\mathbf{x}_l(t_1), \mathbf{x}_l(t_2), \dots, \mathbf{x}_l(t_m)\big)}_{\text{Branch}} \underbrace{N^T_k(\mathbf{\hat{x}},t)}_{\text{Trunk}},
\end{equation}
where $N^B_k$ and $N^T_k$ represent the output of the branch and trunk nets, respectively.
\\

\begin{figure}[htb]
    \centering
    \includegraphics[width=1.1\linewidth]{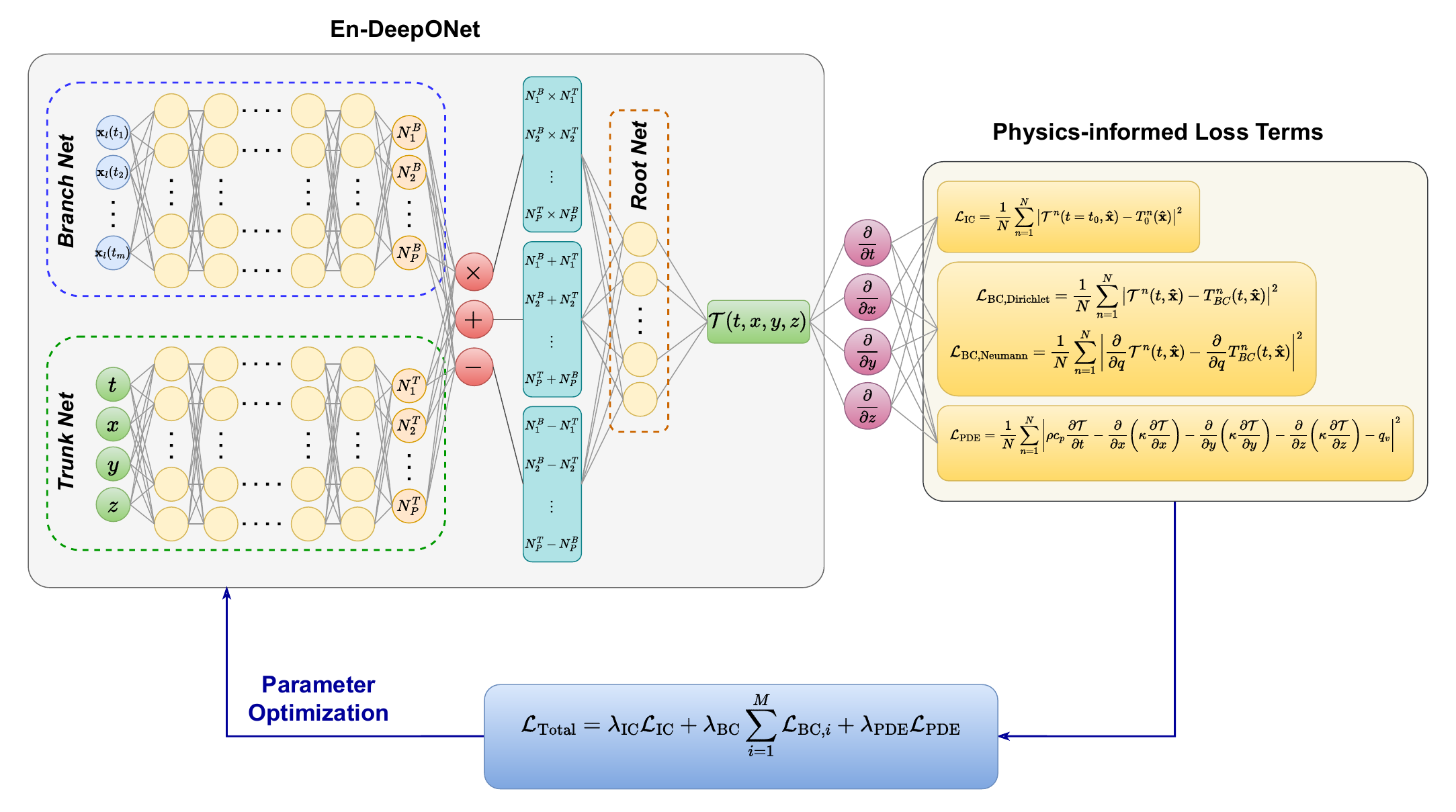}
    \caption{Schematic representation of the En-DeepONet with physics-informed loss functions.}
    \label{fig:pi-deeponet}
\end{figure}

Haghighat et al. \cite{haghighat2024deeponet} have recently found that the standard DeepONet shows severe limitations when applied to moving-solution operators. They proposed the enriched DeepONet (En-DeepONet), that instead of using a simple projection operator (dot product) for evaluating the final output, more operators including summation, subtraction and Hadamard (or element-wise) product are taken into account. This idea is demonstrated in Fig.~\ref{fig:pi-deeponet} and can be mathematically described as,

\begin{equation}
    \label{eqn:en_deeponet}
    \begin{aligned}
        \mathcal{T}(\mathbf{x}_l)(\mathbf{\hat{x}},t) &\approx \sum_{k=1}^{p} w_k^{\times} \Big(N^B_k\big(\mathbf{x}_l(t_1), \mathbf{x}_l(t_2), \dots, \mathbf{x}_l(t_m)\big) \cdot N^T_k(\mathbf{\hat{x}},t) \Big) \\
&+  \sum_{k=1}^{p} w_k^{+} \Big(N^B_k\big(\mathbf{x}_l(t_1), \mathbf{x}_l(t_2), \dots, \mathbf{x}_l(t_m)\big) + N^T_k(\mathbf{\hat{x}},t) \Big) \\
&+  \sum_{k=1}^{p} w_k^{-} \Big(N^B_k\big(\mathbf{x}_l(t_1), \mathbf{x}_l(t_2), \dots, \mathbf{x}_l(t_m)\big) - N^T_k(\mathbf{\hat{x}},t) \Big),
    \end{aligned}
\end{equation}
where each summation ($\sum$) can be considered as a linear layer with the weight factors of $w_k^{\bullet}$ with $\bullet \in \{\cdot,+,-\}$. These series of linear layers are depicted as the so-called root net in En-DeepONet part of Fig.~\ref{fig:pi-deeponet}. Adding physics-informed loss terms (Eqs.~\eqref{eq:loss_ic}-\eqref{eq:loss_pde}), the resulting En-DeepONet is called \pidon{}. This architecture is capable of learning the spatiotemporal evolution of the temperature across multiple tool paths all at once without any labeled data.

We will compare the outputs of the \pidon{} with the PINN counterpart in the following Sections. The presented PINN and \pidon{} share the same loss functions. Moreover, the feed-forward neural network (FNN) of the PINN is comparable with the trunk net in \pidon{} in terms of the number of hidden layers and the activation function. Note that we opt for constant weighting factors as computed by Eq.~\eqref{eq:lambda} and soft boundary conditions as given in Eqs.\eqref{eq:loss_bc_d}-\eqref{eq:loss_bc_n} as their more advanced counterparts (i.e. adaptive weighting factors~\cite{hou2023enhancing} and hard boundary conditions) did not improve the accuracy of the trained networks in our test cases. More information on the hyperparameters of the \pidon{} and PINN is given in Section~\ref{sebsec:criteria}.

\begin{table}[p]
\begin{center}
\caption{Material properties, process parameters, initial and boundary conditions for all test cases.}
\begin{tabular}{ll} 
 \textbf{Material Property} & \textbf{Value} \\ 
 \hline
 \hline
 Density ($\rho$) & 8351.91 $ \text{kg/} \text{m}^3$  \\ 
 Thermal conductivity ($\kappa$) & Eq.~\eqref{eq:prop}  \\ 
 Heat capacity ($c_p$) & Eq.~\eqref{eq:prop} \\ 
 \rule{0pt}{15pt}
 \textbf{Laser Parameter} & \textbf{Value} \\ 
 \hline
 \hline
 Modified laser power ($\eta \times P$) & 150.0 $\text{W}$ \\ 
 Laser beam radius & 450 $\mu\text{m}$ \\ 
 Laser beam scan speed & 0.1 $\text{m/s}$ \\ 
 \rule{0pt}{15pt}
 \textbf{Simulation Condition} & \textbf{Value} \\ 
 \hline
 \hline
 Initial temperature ($T_0$) & 300 $\text{K}$  \\ 
 Ambient temperature ($T_\infty$) & 300 $\text{K}$  \\ 
 Convection heat transfer coefficient ($h_{conv}$) & 10 $\text{W/(} \text{m}^2 \text{K)}$ \\ 
\end{tabular}
\label{tab:laser_parameter}
\end{center}
\end{table}

\subsection{Multi-track scenarios}

\begin{figure}[h]
    \centering
    \begin{subfigure}{0.55\textwidth}
    \includegraphics[width=\textwidth]{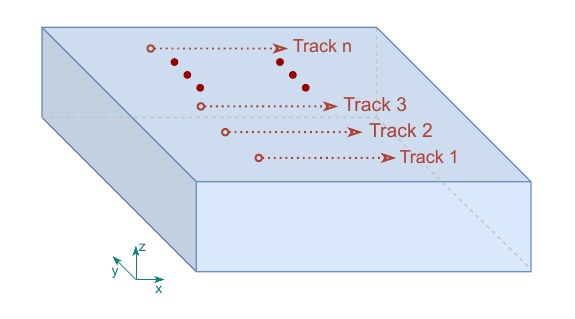}
    \caption{}
    \label{fig:multi_track_a}
    \end{subfigure}
    \begin{subfigure}{0.85\textwidth}
    \includegraphics[width=\textwidth]{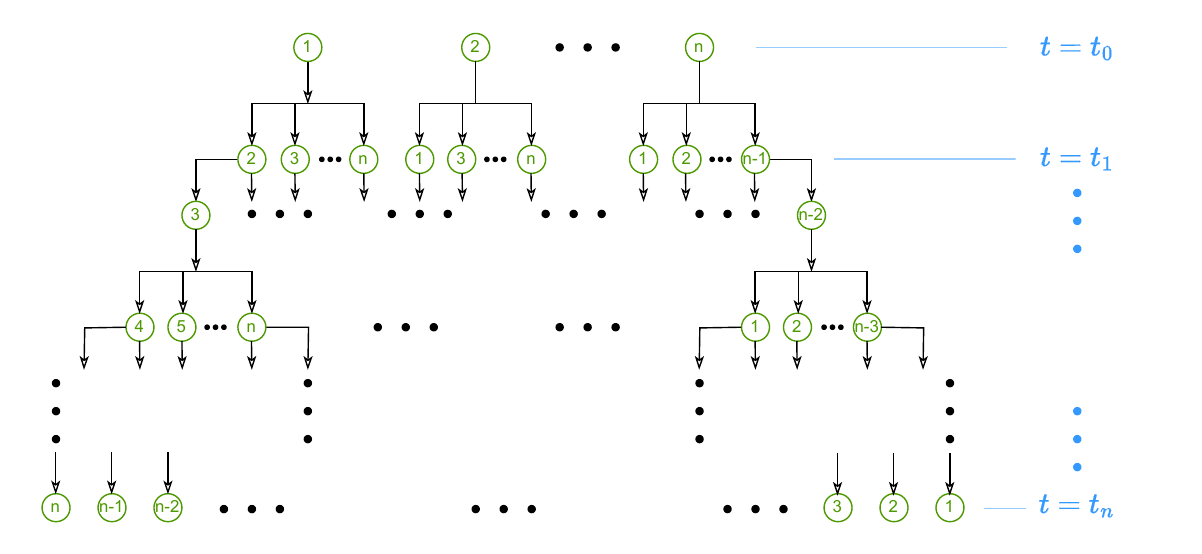}
    \caption{}
    \label{fig:multi_track_b}
    \end{subfigure}   
    \caption{(a) Outline of multiple tracks on a workpiece. (b)Tree structure representation of breaking down multi-track scenarios into $n$ times intervals. Green colored nodes indicate the track numbers.}
    \label{fig:multi_track}
\end{figure}

In this study, we deal with both single-track and multi-track case studies. In multi-track scenarios, the laser successively scans different tracks on the workpiece. In multi-track test cases, a scenario (or a path) is defined as the sequence of tracks. Hence, each scenario can be considered as the combination of multiple tracks. Assuming multi-track manufacturing with $n$ tracks per path as shown in Fig.~\ref{fig:multi_track_a}, the total number of $n!$ different scenarios can be recognized, each consisting of all $n$ tracks, where

\begin{equation}
\begin{aligned}
    n! &= \prod_{i=1}^{n} i \\
     &= n\cdot (n-1) \cdot ... \cdot 3 \cdot 2 \cdot 1.
 \end{aligned}
\end{equation}

In the next Section, we will observe that due to the curse of dimensionality, the \pidon{} faces difficulties when the number of tracks in each scenario increases. To address this challenges, each scenario is broken down into its constituent tracks. We aim to train a dedicated PINN for each track within each scenario. To solve for all possible scenarios, it is essential to identify all potential combinations of tracks that comprise each scenario. To have a better understanding of how individual scenarios are constructed from multiple tracks, we visualize the problem as a tree structure, as shown in Fig.~\ref{fig:multi_track_b}. In this representation, tracks are depicted as nodes, and each scenario is defined  by tracing the branches along the arrows from the top to the bottom nodes. In other words, we split each scenario into $n$ individual stages (or $n$ time intervals,  $T_i=\left[t_{(i-1)}, t_i\right] \, \forall \, i \in \left[1, n\right]$), which correspond to the time needed to process the related tracks. The total number of individual trained PINNs ($N^{\text{train}}$) required for computation of all paths is obtained from
\begin{equation}
    \label{eq:n_train}
    \begin{aligned}
        N^{\text{train}} =  \mathlarger{\sum_{i=0}^{n-1}} \prod_{j=0}^{i} (n-j).
    \end{aligned}
\end{equation}

In a given scenario, each track has both a predecessor and a successor, except for the first and last tracks, which only have a successor or predecessor, respectively. While the boundary conditions remain the same across all tracks, the initial condition of each track is unique and corresponds to the end-state of its predecessor. This ensures a continuous and physically consistent progression of the solution across the sequence of tracks. Once the training of each track is completed, the relevant weights and biases of the network are saved, allowing for efficient storage and retrieval. These trained models can be easily recalled whenever required. This methodology is referred to as Sequential PINN, as the complete solution for a given scenario is constructed by sequentially recalling the trained PINNs for each track within that scenario.


\section{Results and Discussion} \label{sec:result}
\begin{figure}[p]
    \centering
    \includegraphics[width=0.75\textwidth]{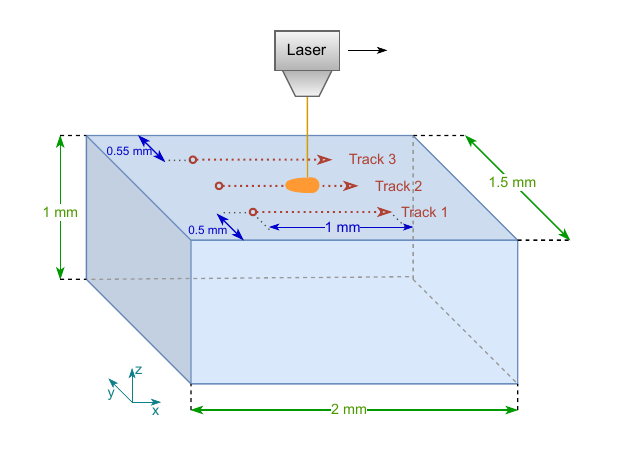}
    \caption{Schematic representation of the computational domain and boundary conditions.}
    \label{fig:domain}
\end{figure}

In this study, we focus on solving the transient heat equation within a three-dimensional computational domain, subject to the boundary conditions illustrated in Fig.~\ref{fig:domain}. The laser can scan along three parallel tracks, allowing for flexible path selection for both single-track and multi-track scenarios. In the single-track case, the laser processes only one of the three tracks at a time. However, in the multi-track scenario, each path is defined as a combination of multiple tracks (i.e. $1\rightarrow2\rightarrow3$, $ 2\rightarrow3\rightarrow1$ etc.). For this specific case which is depicted in Fig.~\ref{fig:domain}, $3!=6$ different paths can be identified, each comprising all three tracks. As a material, we consider Hastelloy X with the following temperature-dependent properties~\cite{hosseini2023single}:
                  
\begin{equation}
\label{eq:prop}
    \begin{aligned}
        \kappa &=  229.87 + 0.0184\,T + 225.10 \tanh \big( 0.0118(T-1816.8) \big)\: \textstyle \frac{W}{m K},\\
        c_p &= 407.62 + 0.142\,T - 61.43 \,\mathrm{e}^{-3.1\times10^{-4}(T-798)^2} + 1054.96 \, \mathrm{e}^{-6.2\times10^{-5}(T-1816.8)^2}\: \textstyle \frac{J}{kg K},
    \end{aligned}
\end{equation}
where $\kappa$ represents the thermal conductivity, $c_p$ denotes the apparent heat capacity, and the temperature ($T$) is expressed in Kelvin. Note that the effect of the latent heat of fusion on the apparent heat capacity has been accounted for in \eqref{eq:prop}$_2$. The process parameters including the material properties, laser parameters, boundary and initial conditions are listed in Table~\ref{tab:laser_parameter}.

\subsection{Evaluation criteria and network parameters} \label{sebsec:criteria}

The performance of the physics-informed surrogate models considered in this study is evaluated using several criteria. Two primary metrics are used to assess accuracy: the relative error and the mean absolute percentage error (MAPE). The relative error is calculated for predicted temperatures and melt pool dimensions, enabling a localized evaluation of the model's ability to capture the steep thermal gradients observed in the melt pool. The MAPE provides an overall measure of accuracy by quantifying the deviation of the predicted temperature from the reference FD solutions across the entire computational domain and is calculated by

\begin{equation}\label{eq:MAPE}
    \text{MAPE} = \frac{1}{n} \sum_{i=1}^{n} \left| \frac{T_{NN,i} - T_{FD,i}}{T_{FD,i}} \right|
\end{equation}
where $T_{NN}$ and $T_{FD}$ represent the \pidon{} (or PINN)  and finite-difference calculated temperatures, while $n$ is the total number of points inside the domain.

The temperature history during the cooling and consolidation phases is also considered to ensure the model's reliability in capturing the temporal evolution of the thermal field. This is crucial for accurately predicting the cooling time and the subsequent effects on material properties and microstructure.

The PI-DeepONet architecture is structured with separate trunk and branch networks. The trunk network consists of 5 hidden layers, each containing 64 neurons, while the branch network has 2 hidden layers, each with 64 neurons. In comparison, the PINN architecture employs a single FNN with 5 hidden layers, each containing 64 neurons. For both PI-DeepONet and PINN, a sine function is used as the activation function in the form of $\sin{\pi \beta x}$, where $\beta$ is an extra trainable parameter. The Limited-memory Broyden–Fletcher–Goldfarb–Shanno (L-BFGS)~\cite{liu1989limited} optimizer is selected for training. L-BFGS is particularly effective for physics-informed neural networks due to its ability to efficiently handle smooth loss landscapes and achieve high precision during convergence~\cite{rathore2024challenges}.


\subsection{Single-track scenarios}

Before analyzing the performance of \pidon{} in multi-track scenarios, we first assess its ability to provide parametric solutions for various tool paths in single-track scenarios. Referring to the domain illustrated in Fig.~\ref{fig:domain}, each single-track scenario involves the laser processing of one out of three possible tracks. In these cases, the laser operates over the same length of the workpiece but starts at different positions. Thus, a single \pidon{} model learns the temperature evolution for three scenarios all at once. For comparison, three separate PINN models are trained individually, with each model focusing on the temperature evolution of a specific scenario.

\input{results/single-tracks}

Fig.~\ref{fig:single-tracks} compares the predicted temperature and meltpool dimensions with the finite-difference solution (as the reference solution). As seen in Fig.~\ref{fig:single-tracks_pinn} and \ref{fig:single-tracks_pideep}, the predicted temperature for both approaches agree well with the reference solution with the MAPE $< 2\%$. However, PINN demonstrates slightly better accuracy in temperature predictions compared to \pidon{}, particularly at higher temperatures. Furthermore, PINN also performs slightly better in terms of average meltpool dimensions as shown in Fig.~\ref{fig:single-tracks_meltpool}. All of the trainings in this study are performed on Nvidia A100 80GB GPU. Table~\ref{tab:single-tracks} shows the total training times and other global metrics of PINN and \pidon{} for single-track scenarios. In estimation of the total training time of the PINN, we assume that only one GPU is available for training. However, in practice, the three different PINNs can be trained in parallel on different GPUs without any overhead. On the other hand, \pidon{} offers a parametric solution that accommodates variations in the tool path, whereas with PINN, we obtain three distinct solutions corresponding to three different tool paths.

\subsection{Multi-tracks scenarios}

In the current test case with 3 parallel tracks ($n=3$), as shown in Fig.~\ref{fig:domain}, the total of 6 (i.e. $3!$) unique paths can be identified. \\

\begin{figure}[p]
    \centering
    \begin{subfigure}{0.48\textwidth}
    \includegraphics[width=\textwidth]{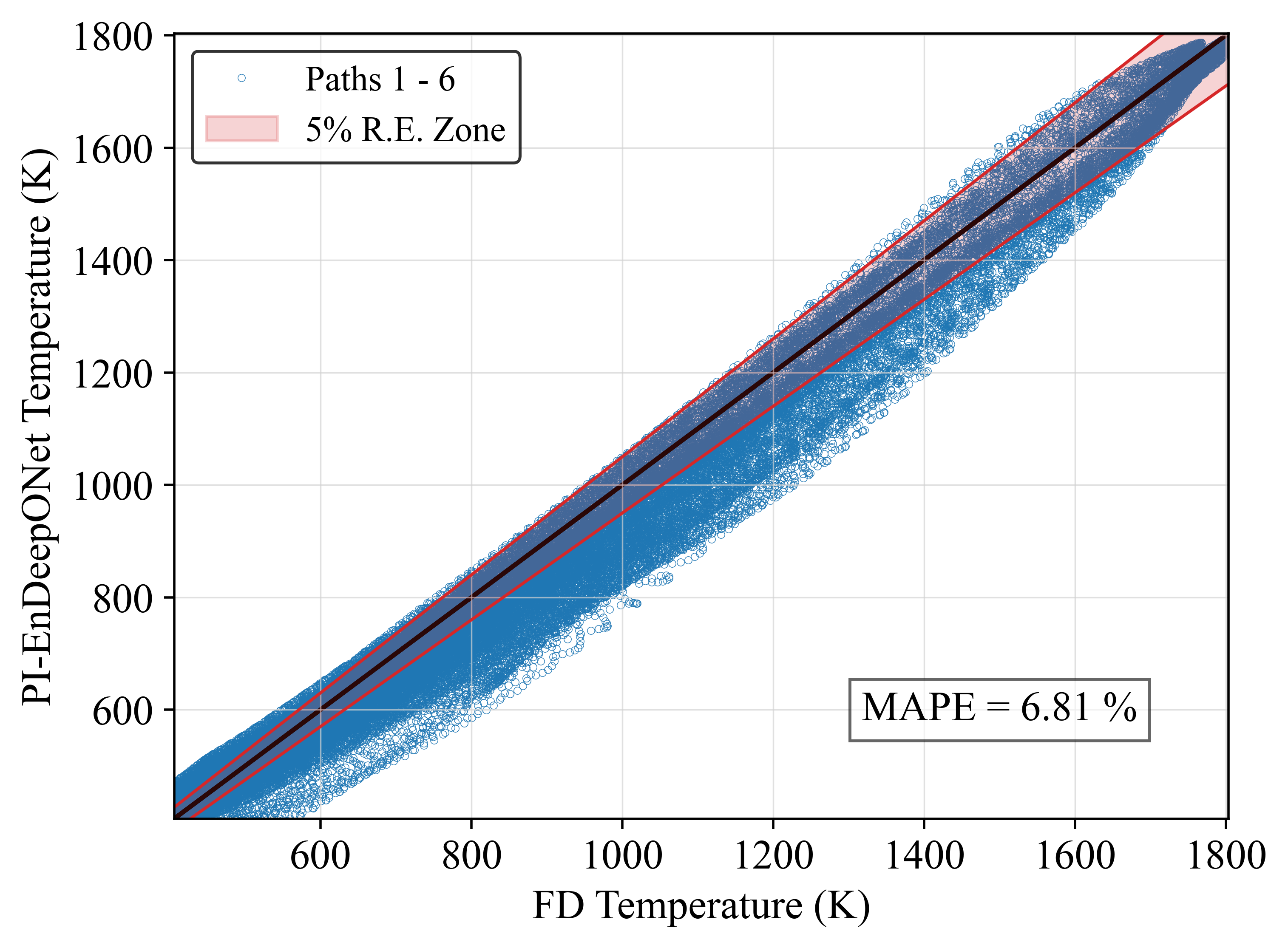}
    \caption{\pidon{}}
    \label{fig:all_path_pideep_a}
    \end{subfigure}
    \begin{subfigure}{0.48\textwidth}
    \includegraphics[width=\textwidth]{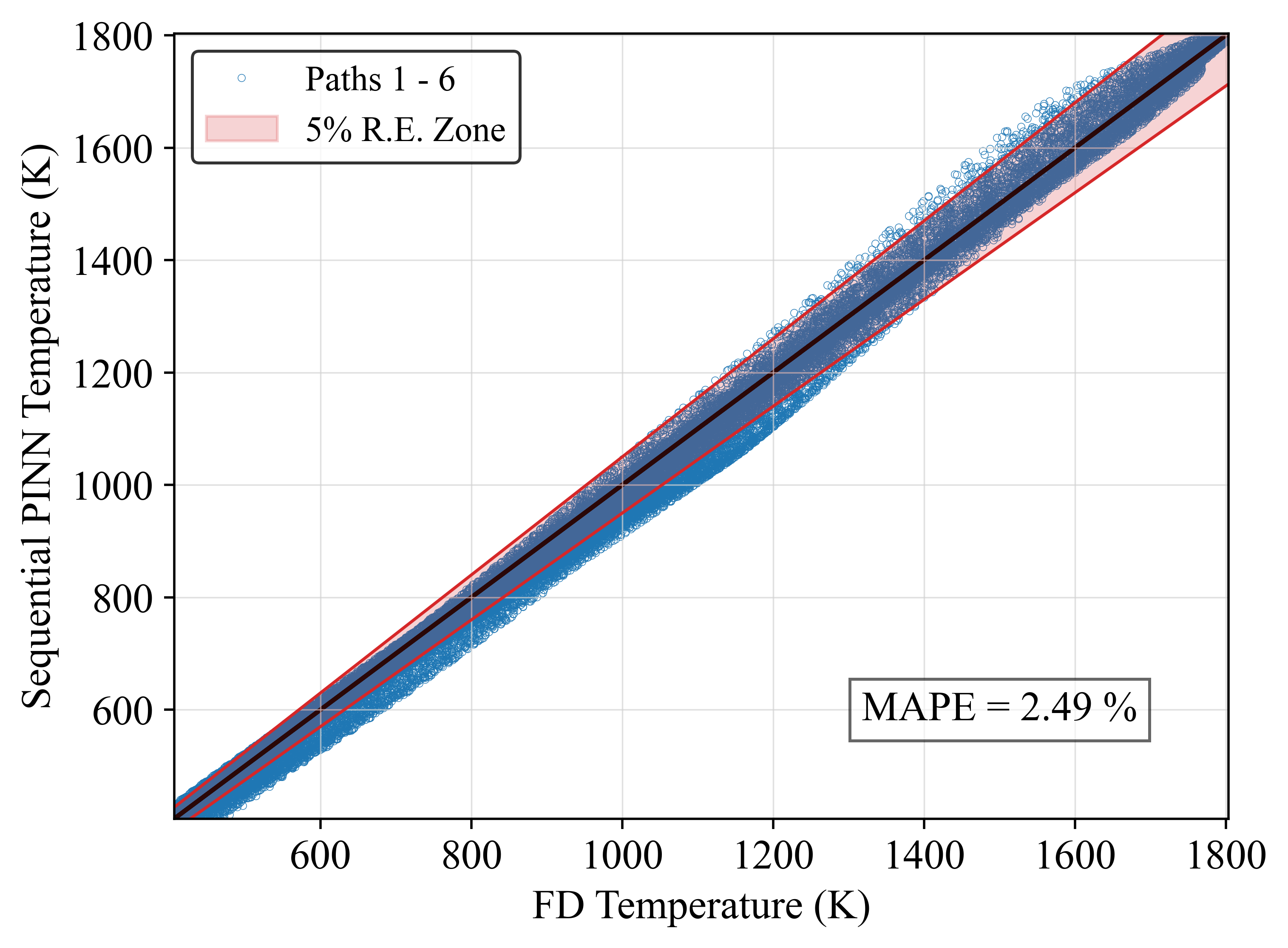}
    \caption{Sequential PINN}
    \label{fig:all_path_pinn_b}
    \end{subfigure}
    \begin{subfigure}{1\textwidth}
    \includegraphics[width=\textwidth]{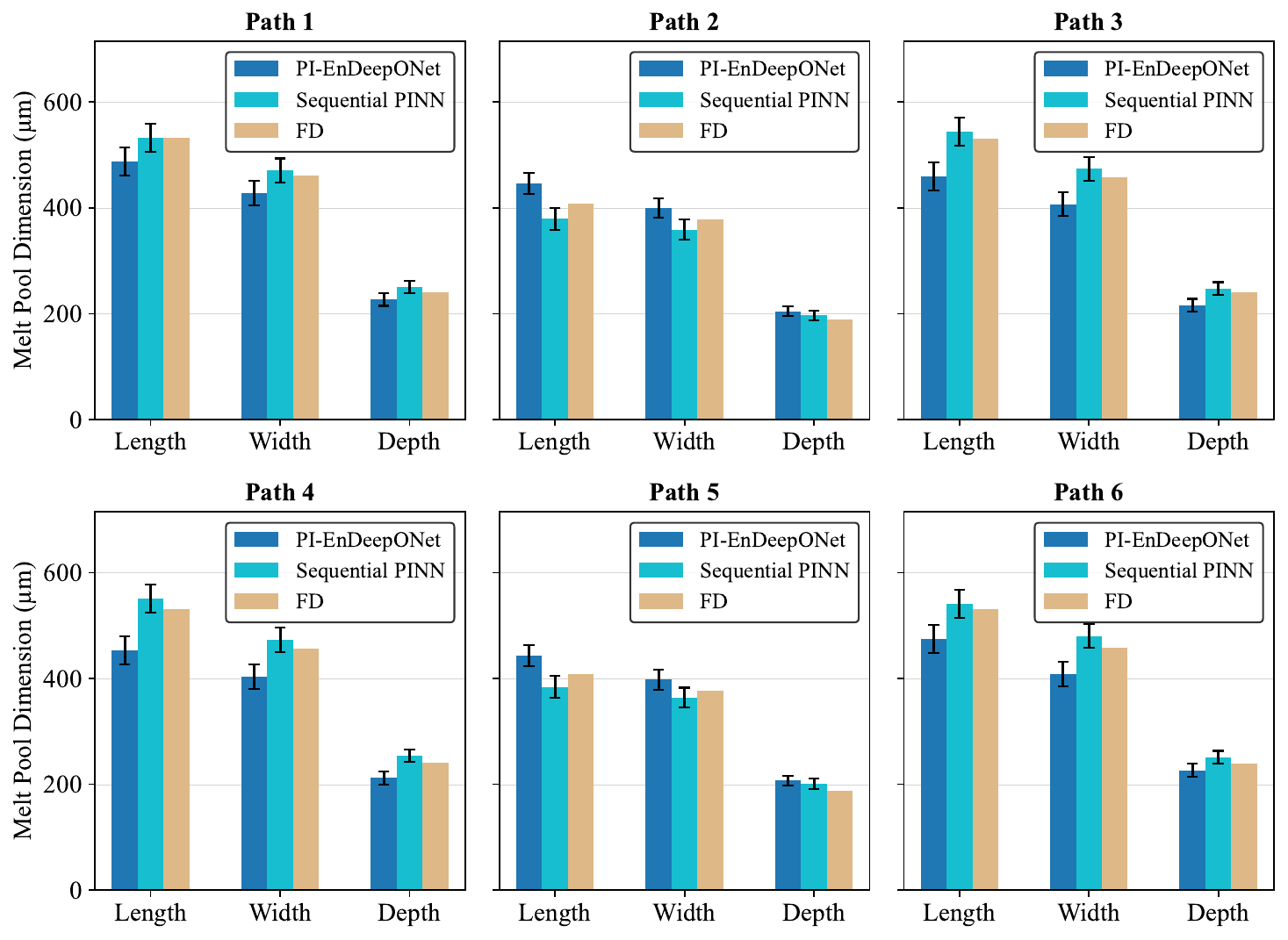}
    \caption{Melt pool dimensions}
    \label{fig:all_path_pideep_pinn_c}
    \end{subfigure}
    \caption{Comparison of the calculated temperature and meltpool dimensions for all multi-track scenarios (paths). R.E. stands for the relative error. (a): \pidon{} solution vs. reference (FD) solution, (b): Sequential PINN solution vs. reference (FD) solution. 5 \% relative error zone is indicated by light red color. (c): Comparison of melt pool dimensions for each method.  Error bars indicate 5\% error with respect to FD solution.}
    \label{fig:multi-tracks}
\end{figure}

\noindent \textbf{\pidon{}:} The \pidon{} is trained on all possible 6 paths, with each path containing 3 tracks. Fig.~\ref{fig:all_path_pideep_a} presents the comparison in the calculated temperature and the melt pool dimensions between the \pidon{} and the reference solution. We experienced the average training time of 17h for all 6 paths. With the $\text{MAPE}=6.81\%$ and the average relative error of 8-10\% in melt pool dimensions, the results clearly indicate that the overall accuracy of the model in multi-track scenarios is not satisfactory.

The \pidon{} model presents an interesting theory and framework for obtaining parametric solutions in terms of the different tool paths. It achieves acceptable accuracy in single-track scenarios. However, as observed, by increasing the number of tracks involved in each scenario the errors are substantially elevated. This can be attributed to the curse of dimensionality. By increasing the number of tracks in each scenario, the computational complexity across the domain increases. In single-track scenarios, the tool path moves only along a 1-dimensional track. However, in multi-track scenarios, the tool path moves along multiple tracks over a surface. This increase in the dimensionality of the heat source movement introduces challenges in capturing the complex spatiotemporal dynamics of heat transfer, where interactions between tracks create localized reheating and sharper thermal gradients. These coupled interactions demand high spatial and temporal collocation points, significantly increasing the complexity of the problem. Additionally, the growing input function (the input to the branch net) dimensionality amplifies the computational cost and training time, making it harder to maintain high accuracy and low error rates across all scenarios. To overcome these challenges and improve the accuracy, the next section introduces a sequential approach for obtaining the solutions of multi-track scenarios. This method leverages PINNs to achieve a balance between acceptable accuracy and reasonable training times.\\



\noindent \textbf{Sequential PINN:} For the workpiece which undergoes 3 parallel tracks ($n=3$),  15 individual PINNs ($N^{train}=15$ according to Eq.~\ref{eq:n_train}) are required to obtain the whole solution across all 6 scenarios. To evaluate the accuracy of the sequential PINN approach, the predicted temperature as well as the meltpool dimensions are compared with the reference solution in Fig.~\ref{fig:all_path_pinn_b} and Fig.~\ref{fig:all_path_pideep_pinn_c}. As observed in Fig.~\ref{fig:all_path_pinn_b}, the consecutive PINN model is capable of approximating the reference solutions for 6 scenarios. Thanks to the adaptive clustering of collocation points in the vicinity of the laser spot, at temperatures higher than 1600 K, where the melt pool is developed, the relative errors lie within the 5\% error with respect to the reference solutions. As indicated, the MAPE for all the points inside the domain is less than 2.5\% which confirms the consistency of the sequential PINN model in computing the temperature field for multi-track scenarios. Successful clustering also helps obtaining relatively acceptable accuracies (less than 5 \%) in estimating the melt pool dimensions as shown in Fig.~\ref{fig:all_path_pideep_pinn_c}. To gain deeper insight into the distribution of relative errors across the domain, Fig.~\ref{fig:all_compare_T} illustrates the temperature contours for PINN and finite-difference solutions in both $xy-$ and $xz-$ planes along the laser direction at the middle of the process ($t=15 \: \text{ms}$), together with the distribution of relative error. The temperature distributions demonstrate very good agreement with the reference solution throughout the entire domain and across all scenarios. Table~\ref{tab:multi-tracks} provides a comparison of the training times and other global metrics between the sequential PINN and \pidon{}. The time required to solve all scenarios using the sequential PINN is approximately 8.5 times less than that of \pidon{}. The above evaluations highlight the effectiveness of the sequential PINN approach as a reliable surrogate for efficient and accurate computation of temperature distributions in multi-track scenarios during the melting phase.

\begin{table}[tb]
    \centering
    \caption{Comparison of the training times an other metrics for multi-track scenarios between sequential PINN and \pidon{}.}
    \begin{tabular}{c|c|c}
                    & Sequential PINN & \pidon{} \\
                    \hline
                    \hline
     Total training time & 2 h & 17 h \\
     MAPE in temperature & 2.49 \% & 6.81 \% \\
     Average relative error in melt pool length & 3.58 \% & 10.80 \% \\
     Average relative error in melt pool width & 3.81 \% & 8.62 \% \\
     Average relative error in melt pool depth & 4.89 \% & 8.53 \% 
    \end{tabular}
    \label{tab:multi-tracks}
\end{table}

\begin{figure}[h]
    \centering
    \begin{subfigure}{0.85\textwidth}
    \includegraphics[width=\textwidth]{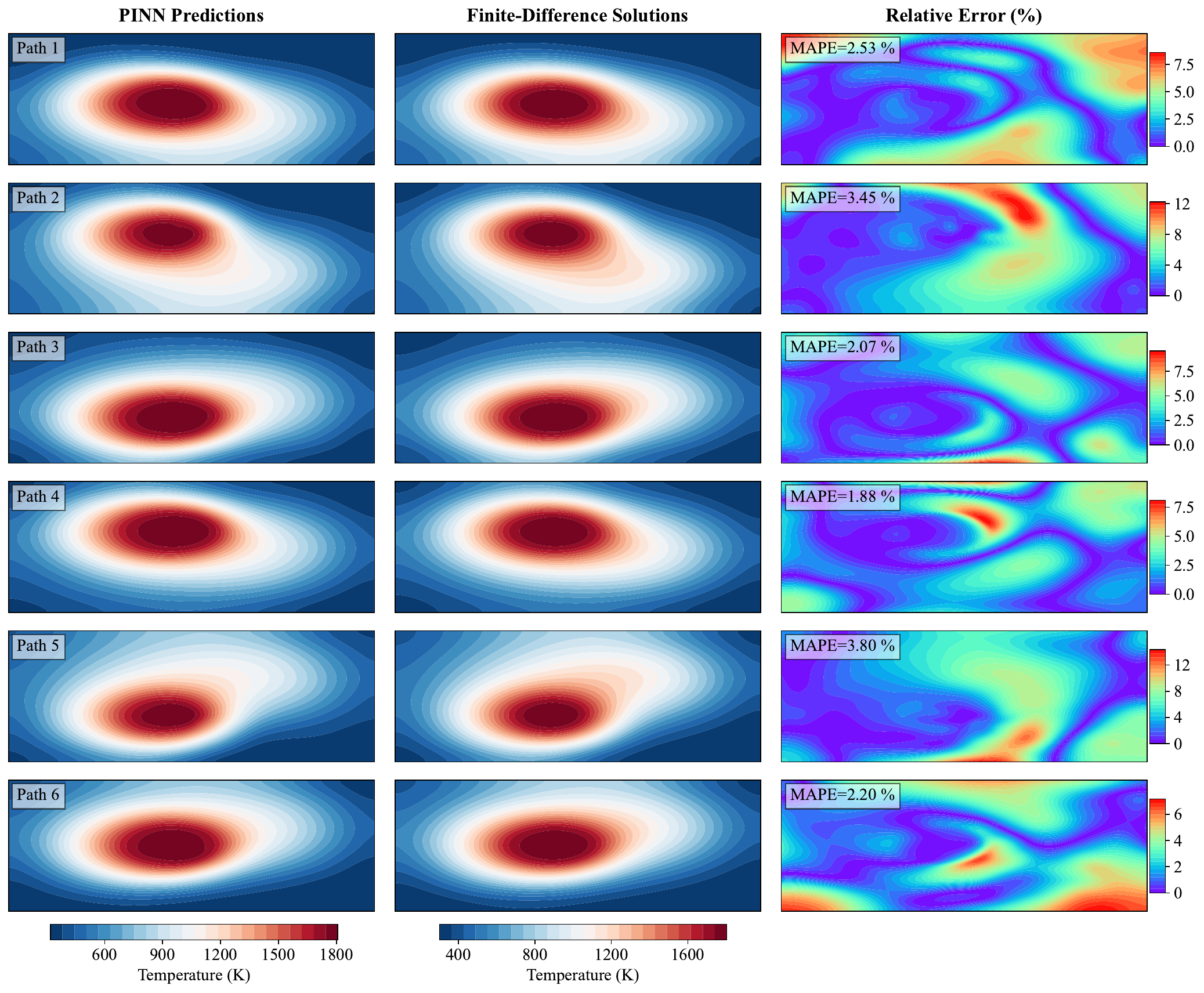}
    \caption{$xy$-plane}
    \label{fig:multi_a}
    \end{subfigure}
    \begin{subfigure}{0.85\textwidth}
    \includegraphics[width=\textwidth]{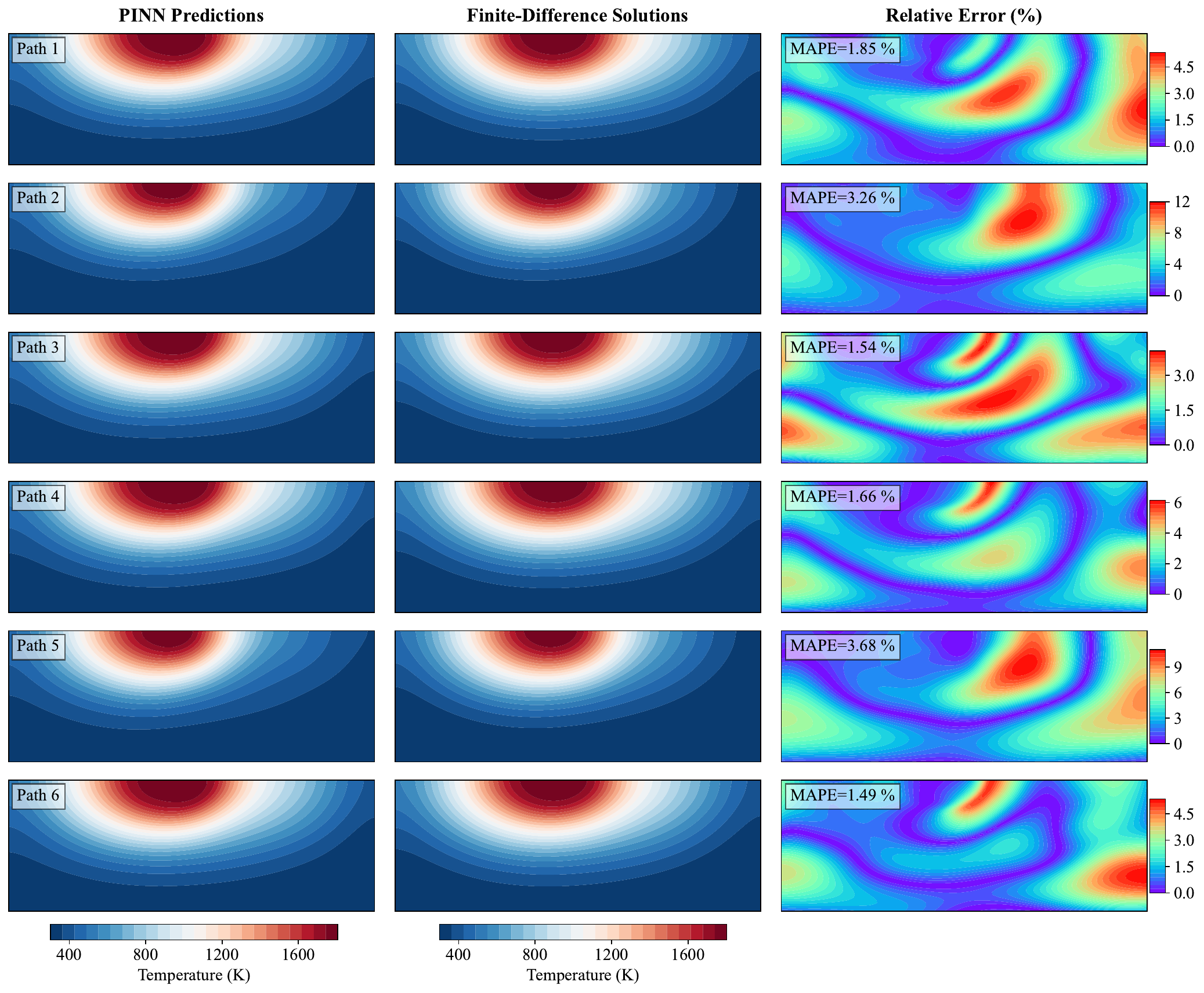}
    \caption{$xz$-plane}
    \label{fig:multi_b}
    \end{subfigure}
    
    \caption{Comparison between the PINN predictions and finite-difference solutions for all paths in both $xy-$ and $xz-$ planes at $t=15 \: \text{ms}$. The $xy$ plane is the top view in Fig.~\ref{fig:domain} and the $xz$ plane is a cut along the laser path.}
    \label{fig:all_compare_T}
\end{figure}

\begin{figure}[h]
    \centering
    \begin{subfigure}{0.65\textwidth}
    \includegraphics[width=\textwidth]{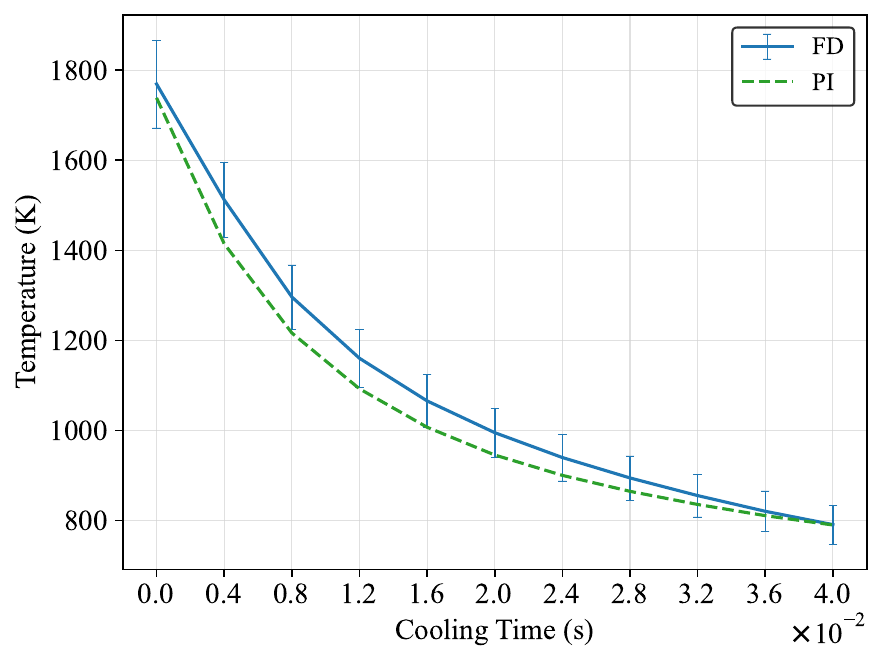}
    \caption{}
    \label{fig:cooling}
    \end{subfigure}
    \begin{subfigure}{0.65\textwidth}
    \includegraphics[width=\textwidth]{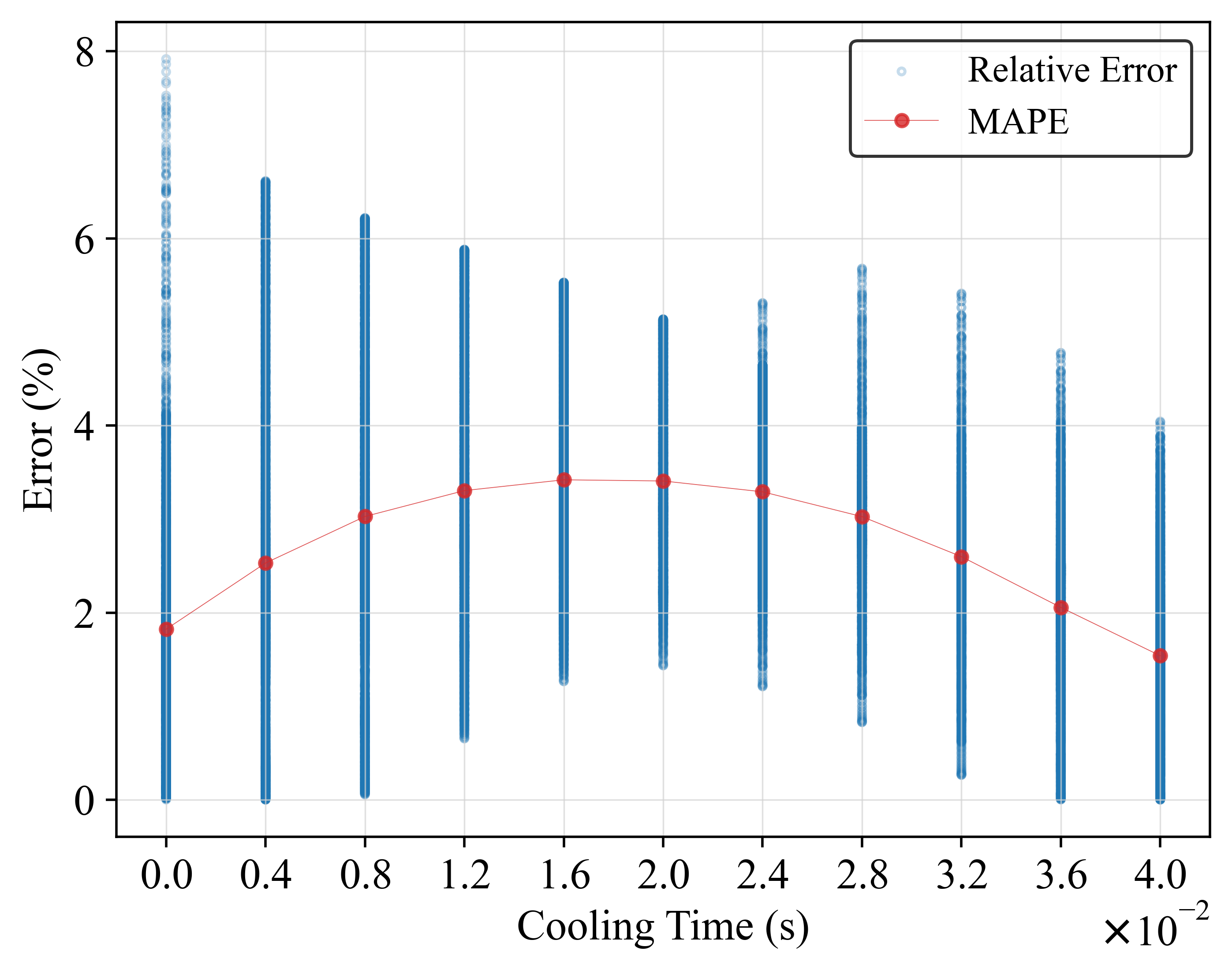}
    \caption{}
    \label{fig:cooling_all}
    \end{subfigure}
    
    \caption{(a) Comparison of the temperature history at the center of the meltpool during the cooling and consolidation phase. Error bars represent 5\% error with respect to FD solution. (b) The relative errors in calculated temperature at different positions across the whole computational domain during the cooling and consolidation phases for all paths. Each position is represented by a blue dot in the plot. Darker color implies aggregation of the relative error values. The MAPE at each time step is shown by the red dot.}
    \label{fig:all_compare_T}
\end{figure}

Besides acceptable accuracy in temperature distribution and meltpool dimensions, a successful surrogate should be able of providing a consistent temperature history during the consolidation phase. Influencing the microstructure development, the consolidation affects the mechanical properties and overall quality of the manufactured item. To evaluate the accuracy of temperature predictions during cooling and consolidation, we compare the temperature history at various locations within the melt pool against the reference solution across all scenarios. Fig.~\ref{fig:cooling} shows the evolution of the temperature at the center of the melt pool over time and compares it with the reference solution for the scenario number 1. Similar trends are also observed for different locations throughout the melt pool across all scenarios. The relative errors in the temperature history profiles across the melt pool region remain below 5\% for all scenarios, ensuring a high level of agreement with the reference solution.

To provide a more comprehensive assessment of the model's consistency, the distribution of relative errors and the MAPE in the computed temperatures at various process times for all data points across the domain in all six scenarios are illustrated in Fig.~\ref{fig:cooling_all}. The MAPE, consistently remaining below 5\%, confirms the reliability of the developed surrogate model in accurately evaluating the cooling and the consolidation phases.

    


\section{Conclusion and Outlook}\label{sec:conclusion}
This study explored the application of physics-informed neural networks (PINNs) and Physics-Informed neural operators (PI-EnDeepONet) for solving the 3-dimensional transient heat equation in laser powder bed fusion (LPBF) processes. Our primary focus was to develop parametric surrogate models for predicting the temperature field in single-track and multi-track scenarios. While PI-EnDeepONet demonstrated success in obtaining parametric solutions for single-track scenarios with high accuracy ($\text{MAPE} < 2\%$), its performance significantly deteriorated in multi-track cases due to the curse of dimensionality and increased computational complexity. In contrast, the sequential PINN approach proved to be an effective and scalable alternative. By training individual PINNs for each track and recalling them sequentially, this method maintained low errors ($\text{MAPE} < 2.5\%$) and accurate melt pool predictions, while also reducing training time by a factor of 8.5 compared to PI-EnDeepONet. Moreover, the sequential PINN effectively captured the cooling phase, ensuring accurate temperature evolution throughout the entire process, which is critical for microstructural predictions in AM. This offers significant potential for rapid process optimization and real-time monitoring in additive manufacturing, particularly for tool path planning where high-fidelity simulations are often computationally prohibitive.

Despite these promising results, several challenges remain. One key limitation of the sequential PINN approach is the potential accumulation of errors over multiple tracks, which could impact predictions in larger-scale multi-track or multi-layer simulations. Future research should explore methods to mitigate these errors. Extending this framework to multi-layer AM processes would also be a next step, given the increased complexity of thermal interactions.

\section*{Acknowledgement}

HW acknowledges the DFG for funding within the individual research grant "Data-driven simulation of microstructure in powder bed fusion processes" with project ID 512730472. HS acknowledges helpful discussions and technical support of David Anton in this study.

\appendix 


\bibliographystyle{plainnat}
\bibliography{literature}

\end{document}

%% file: definitions.tex

\newcommand{\pidon}{PI-EnDeepONet}
\newcommand{\HS}{\textcolor{orange}}
\newcommand{\HW}{\textcolor{blue}}

\newcommand{\lrap}[1]{``#1''}

\newcommand{\Caption}[1]{ \begin{singlespace} \em{\caption{#1}} \end{singlespace} }

\long\def\symbolfootnote[#1]#2{\begingroup \def\thefootnote{\fnsymbol{footnote}}\footnote[#1]{#2} \endgroup} 

\renewcommand{\vec}[1]{ \ensuremath{ \mathbf{ #1 } } }
\newcommand{\ten}[1]{ \ensuremath {\mathbf{#1} } }

\newcommand{\gvec}[1]{ \ensuremath{ \boldsymbol{ #1 } } }
\newcommand{\gten}[1]{ \ensuremath {\boldsymbol{#1} } }


\newcommand{\mn}{_{\mbox{\tiny{N}}}}
\newcommand{\mt}{_{\mbox{\tiny{T}}}}

\newcommand{\dint}{\mbox{d}}
\newcommand{\de}{\,\mbox{det}\,}
\newcommand{\gra}{\,\mbox{grad}\,}
\newcommand{\Gra}{\,\mbox{Grad}\,}
\newcommand{\tra}{\,\mbox{tr}\,}
\newcommand{\Div}{\mbox{Div}\,}
\renewcommand{\div}{\mbox{div}\,}
\newcommand{\sign}{\mbox{sign}}

\newcommand{\ncdot}{\hspace{-0.14cm}\cdot}

\newcommand{\hata}{\ensuremath{ \widehat{a} }}
\newcommand{\hatb}{\ensuremath{ \widehat{b} }}
\newcommand{\hatc}{\ensuremath{ \widehat{c} }}
\newcommand{\hatd}{\ensuremath{ \widehat{d} }}
\newcommand{\hate}{\ensuremath{ \widehat{e} }}
\newcommand{\hatf}{\ensuremath{ \widehat{f} }}
\newcommand{\hatg}{\ensuremath{ \widehat{g} }}
\newcommand{\hath}{\ensuremath{ \widehat{h} }}
\newcommand{\hati}{\ensuremath{ \widehat{i} }}
\newcommand{\hatj}{\ensuremath{ \widehat{j} }}
\newcommand{\hatk}{\ensuremath{ \widehat{k} }}
\newcommand{\hatl}{\ensuremath{ \widehat{l} }}
\newcommand{\hatm}{\ensuremath{ \widehat{m} }}
\newcommand{\hatn}{\ensuremath{ \widehat{n} }}
\newcommand{\hato}{\ensuremath{ \widehat{o} }}
\newcommand{\hatp}{\ensuremath{ \widehat{p} }}
\newcommand{\hatq}{\ensuremath{ \widehat{q} }}
\newcommand{\hatr}{\ensuremath{ \widehat{r} }}
\newcommand{\hats}{\ensuremath{ \widehat{s} }}
\newcommand{\hatt}{\ensuremath{ \widehat{t} }}
\newcommand{\hatu}{\ensuremath{ \widehat{u} }}
\newcommand{\hatv}{\ensuremath{ \widehat{v} }}
\newcommand{\hatw}{\ensuremath{ \widehat{w} }}
\newcommand{\hatx}{\ensuremath{ \widehat{x} }}
\newcommand{\haty}{\ensuremath{ \widehat{y} }}
\newcommand{\hatz}{\ensuremath{ \widehat{z} }}

\newcommand{\hatA}{\ensuremath{ \widehat{A} }}
\newcommand{\hatB}{\ensuremath{ \widehat{B} }}
\newcommand{\hatC}{\ensuremath{ \widehat{C} }}
\newcommand{\hatD}{\ensuremath{ \widehat{D} }}
\newcommand{\hatE}{\ensuremath{ \widehat{E} }}
\newcommand{\hatF}{\ensuremath{ \widehat{F} }}
\newcommand{\hatG}{\ensuremath{ \widehat{G} }}
\newcommand{\hatH}{\ensuremath{ \widehat{H} }}
\newcommand{\hatI}{\ensuremath{ \widehat{I} }}
\newcommand{\hatJ}{\ensuremath{ \widehat{J} }}
\newcommand{\hatK}{\ensuremath{ \widehat{K} }}
\newcommand{\hatL}{\ensuremath{ \widehat{L} }}
\newcommand{\hatM}{\ensuremath{ \widehat{M} }}
\newcommand{\hatN}{\ensuremath{ \widehat{N} }}
\newcommand{\hatO}{\ensuremath{ \widehat{O} }}
\newcommand{\hatP}{\ensuremath{ \widehat{P} }}
\newcommand{\hatQ}{\ensuremath{ \widehat{Q} }}
\newcommand{\hatR}{\ensuremath{ \widehat{R} }}
\newcommand{\hatS}{\ensuremath{ \widehat{S} }}
\newcommand{\hatT}{\ensuremath{ \widehat{T} }}
\newcommand{\hatU}{\ensuremath{ \widehat{U} }}
\newcommand{\hatV}{\ensuremath{ \widehat{V} }}
\newcommand{\hatW}{\ensuremath{ \widehat{W} }}
\newcommand{\hatX}{\ensuremath{ \widehat{X} }}
\newcommand{\hatY}{\ensuremath{ \widehat{Y} }}
\newcommand{\hatZ}{\ensuremath{ \widehat{Z }}}

\newcommand{\tila}{\ensuremath{ \widetilde{a} }}
\newcommand{\tilb}{\ensuremath{ \widetilde{b} }}
\newcommand{\tilc}{\ensuremath{ \widetilde{c} }}
\newcommand{\tild}{\ensuremath{ \widetilde{d} }}
\newcommand{\tile}{\ensuremath{ \widetilde{e} }}
\newcommand{\tilf}{\ensuremath{ \widetilde{f} }}
\newcommand{\tilg}{\ensuremath{ \widetilde{g} }}
\newcommand{\tilh}{\ensuremath{ \widetilde{h} }}
\newcommand{\tili}{\ensuremath{ \widetilde{i} }}
\newcommand{\tilj}{\ensuremath{ \widetilde{j} }}
\newcommand{\tilk}{\ensuremath{ \widetilde{k} }}
\newcommand{\till}{\ensuremath{ \widetilde{l} }}
\newcommand{\tilm}{\ensuremath{ \widetilde{m} }}
\newcommand{\tiln}{\ensuremath{ \widetilde{n} }}
\newcommand{\tilo}{\ensuremath{ \widetilde{o} }}
\newcommand{\tilp}{\ensuremath{ \widetilde{p} }}
\newcommand{\tilq}{\ensuremath{ \widetilde{q} }}
\newcommand{\tilr}{\ensuremath{ \widetilde{r} }}
\newcommand{\tils}{\ensuremath{ \widetilde{s} }}
\newcommand{\tilt}{\ensuremath{ \widetilde{t} }}
\newcommand{\tilu}{\ensuremath{ \widetilde{u} }}
\newcommand{\tilv}{\ensuremath{ \widetilde{v} }}
\newcommand{\tilw}{\ensuremath{ \widetilde{w} }}
\newcommand{\tilx}{\ensuremath{ \widetilde{x} }}
\newcommand{\tily}{\ensuremath{ \widetilde{y} }}
\newcommand{\tilz}{\ensuremath{ \widetilde{z} }}

\newcommand{\tilA}{\ensuremath{ \widetilde{A} }}
\newcommand{\tilB}{\ensuremath{ \widetilde{B} }}
\newcommand{\tilC}{\ensuremath{ \widetilde{C} }}
\newcommand{\tilD}{\ensuremath{ \widetilde{D} }}
\newcommand{\tilE}{\ensuremath{ \widetilde{E} }}
\newcommand{\tilF}{\ensuremath{ \widetilde{F} }}
\newcommand{\tilG}{\ensuremath{ \widetilde{G} }}
\newcommand{\tilH}{\ensuremath{ \widetilde{H} }}
\newcommand{\tilI}{\ensuremath{ \widetilde{I} }}
\newcommand{\tilJ}{\ensuremath{ \widetilde{J} }}
\newcommand{\tilK}{\ensuremath{ \widetilde{K} }}
\newcommand{\tilL}{\ensuremath{ \widetilde{L} }}
\newcommand{\tilM}{\ensuremath{ \widetilde{M} }}
\newcommand{\tilN}{\ensuremath{ \widetilde{N} }}
\newcommand{\tilO}{\ensuremath{ \widetilde{O} }}
\newcommand{\tilP}{\ensuremath{ \widetilde{P} }}
\newcommand{\tilQ}{\ensuremath{ \widetilde{Q} }}
\newcommand{\tilR}{\ensuremath{ \widetilde{R} }}
\newcommand{\tilS}{\ensuremath{ \widetilde{S} }}
\newcommand{\tilT}{\ensuremath{ \widetilde{T} }}
\newcommand{\tilU}{\ensuremath{ \widetilde{U} }}
\newcommand{\tilV}{\ensuremath{ \widetilde{V} }}
\newcommand{\tilW}{\ensuremath{ \widetilde{W} }}
\newcommand{\tilX}{\ensuremath{ \widetilde{X} }}
\newcommand{\tilY}{\ensuremath{ \widetilde{Y} }}
\newcommand{\tilZ}{\ensuremath{ \widetilde{Z }}}

\newcommand{\calA}{\ensuremath{ \mathcal{A} }}
\newcommand{\calB}{\ensuremath{ \mathcal{B} }}
\newcommand{\calC}{\ensuremath{ \mathcal{C} }}
\newcommand{\calD}{\ensuremath{ \mathcal{D} }}
\newcommand{\calE}{\ensuremath{ \mathcal{E} }}
\newcommand{\calF}{\ensuremath{ \mathcal{F} }}
\newcommand{\calG}{\ensuremath{ \mathcal{G} }}
\newcommand{\calH}{\ensuremath{ \mathcal{H} }}
\newcommand{\calI}{\ensuremath{ \mathcal{I} }}
\newcommand{\calJ}{\ensuremath{ \mathcal{J} }}
\newcommand{\calK}{\ensuremath{ \mathcal{K} }}
\newcommand{\calL}{\ensuremath{ \mathcal{L} }}
\newcommand{\calM}{\ensuremath{ \mathcal{M} }}
\newcommand{\calN}{\ensuremath{ \mathcal{N} }}
\newcommand{\calO}{\ensuremath{ \mathcal{O} }}
\newcommand{\calP}{\ensuremath{ \mathcal{P} }}
\newcommand{\calQ}{\ensuremath{ \mathcal{Q} }}
\newcommand{\calR}{\ensuremath{ \mathcal{R} }}
\newcommand{\calS}{\ensuremath{ \mathcal{S} }}
\newcommand{\calT}{\ensuremath{ \mathcal{T} }}
\newcommand{\calU}{\ensuremath{ \mathcal{U} }}
\newcommand{\calV}{\ensuremath{ \mathcal{V} }}
\newcommand{\calW}{\ensuremath{ \mathcal{W} }}
\newcommand{\calX}{\ensuremath{ \mathcal{X} }}
\newcommand{\calY}{\ensuremath{ \mathcal{Y} }}
\newcommand{\calZ}{\ensuremath{ \mathcal{Z} }}

\newcommand{\bbA}{\ensuremath{ \mathbb{A} }}
\newcommand{\bbB}{\ensuremath{ \mathbb{B} }}
\newcommand{\bbC}{\ensuremath{ \mathbb{C} }}
\newcommand{\bbD}{\ensuremath{ \mathbb{D} }}
\newcommand{\bbE}{\ensuremath{ \mathbb{E} }}
\newcommand{\bbF}{\ensuremath{ \mathbb{F} }}
\newcommand{\bbG}{\ensuremath{ \mathbb{G} }}
\newcommand{\bbH}{\ensuremath{ \mathbb{H} }}
\newcommand{\bbI}{\ensuremath{ \mathbb{I} }}
\newcommand{\bbJ}{\ensuremath{ \mathbb{J} }}
\newcommand{\bbK}{\ensuremath{ \mathbb{K} }}
\newcommand{\bbL}{\ensuremath{ \mathbb{L} }}
\newcommand{\bbM}{\ensuremath{ \mathbb{M} }}
\newcommand{\bbN}{\ensuremath{ \mathbb{N} }}
\newcommand{\bbO}{\ensuremath{ \mathbb{O} }}
\newcommand{\bbP}{\ensuremath{ \mathbb{P} }}
\newcommand{\bbQ}{\ensuremath{ \mathbb{Q} }}
\newcommand{\bbR}{\ensuremath{ \mathbb{R} }}
\newcommand{\bbS}{\ensuremath{ \mathbb{S} }}
\newcommand{\bbT}{\ensuremath{ \mathbb{T} }}
\newcommand{\bbU}{\ensuremath{ \mathbb{U} }}
\newcommand{\bbV}{\ensuremath{ \mathbb{V} }}
\newcommand{\bbW}{\ensuremath{ \mathbb{W} }}
\newcommand{\bbX}{\ensuremath{ \mathbb{X} }}
\newcommand{\bbY}{\ensuremath{ \mathbb{Y} }}
\newcommand{\bbZ}{\ensuremath{ \mathbb{Z} }}

\newcommand{\tenA}{\ensuremath{ \ten{A} }}
\newcommand{\tenB}{\ensuremath{ \ten{B} }}
\newcommand{\tenC}{\ensuremath{ \ten{C} }}
\newcommand{\tenD}{\ensuremath{ \ten{D} }}
\newcommand{\tenE}{\ensuremath{ \ten{E} }}
\newcommand{\tenF}{\ensuremath{ \ten{F} }}
\newcommand{\tenG}{\ensuremath{ \ten{G} }}
\newcommand{\tenH}{\ensuremath{ \ten{H} }}
\newcommand{\tenI}{\ensuremath{ \ten{I} }}
\newcommand{\tenJ}{\ensuremath{ \ten{J} }}
\newcommand{\tenK}{\ensuremath{ \ten{K} }}
\newcommand{\tenL}{\ensuremath{ \ten{L} }}
\newcommand{\tenM}{\ensuremath{ \ten{M} }}
\newcommand{\tenN}{\ensuremath{ \ten{N} }}
\newcommand{\tenO}{\ensuremath{ \ten{O} }}
\newcommand{\tenP}{\ensuremath{ \ten{P} }}
\newcommand{\tenQ}{\ensuremath{ \ten{Q} }}
\newcommand{\tenR}{\ensuremath{ \ten{R} }}
\newcommand{\tenS}{\ensuremath{ \ten{S} }}
\newcommand{\tenT}{\ensuremath{ \ten{T} }}
\newcommand{\tenU}{\ensuremath{ \ten{U} }}
\newcommand{\tenV}{\ensuremath{ \ten{V} }}
\newcommand{\tenW}{\ensuremath{ \ten{W} }}
\newcommand{\tenX}{\ensuremath{ \ten{X} }}
\newcommand{\tenY}{\ensuremath{ \ten{Y} }}
\newcommand{\tenZ}{\ensuremath{ \ten{Z} }}

\newcommand{\tena}{\ensuremath{ \ten{a} }}
\newcommand{\tenb}{\ensuremath{ \ten{b} }}
\newcommand{\tenc}{\ensuremath{ \ten{c} }}
\newcommand{\tend}{\ensuremath{ \ten{d} }}
\newcommand{\tene}{\ensuremath{ \ten{e} }}
\newcommand{\tenf}{\ensuremath{ \ten{f} }}
\newcommand{\teng}{\ensuremath{ \ten{g} }}
\newcommand{\tenh}{\ensuremath{ \ten{h} }}
\newcommand{\teni}{\ensuremath{ \ten{i} }}
\newcommand{\tenj}{\ensuremath{ \ten{j} }}
\newcommand{\tenk}{\ensuremath{ \ten{k} }}
\newcommand{\tenl}{\ensuremath{ \ten{l} }}
\newcommand{\tenm}{\ensuremath{ \ten{m} }}
\newcommand{\tenn}{\ensuremath{ \ten{n} }}
\newcommand{\teno}{\ensuremath{ \ten{o} }}
\newcommand{\tenp}{\ensuremath{ \ten{p} }}
\newcommand{\tenq}{\ensuremath{ \ten{q} }}
\newcommand{\tenr}{\ensuremath{ \ten{r} }}
\newcommand{\tens}{\ensuremath{ \ten{s} }}
\newcommand{\tent}{\ensuremath{ \ten{t} }}
\newcommand{\tenu}{\ensuremath{ \ten{u} }}
\newcommand{\tenv}{\ensuremath{ \ten{v} }}
\newcommand{\tenw}{\ensuremath{ \ten{w} }}
\newcommand{\tenx}{\ensuremath{ \ten{x} }}
\newcommand{\teny}{\ensuremath{ \ten{y} }}
\newcommand{\tenz}{\ensuremath{ \ten{z} }}

\newcommand{\ttena}{\dot{\tena}}
\newcommand{\ttenb}{\dot{\tenb}}
\newcommand{\ttenc}{\dot{\tenc}}
\newcommand{\ttend}{\dot{\tend}}
\newcommand{\ttene}{\dot{\tene}}
\newcommand{\ttenf}{\dot{\tenf}}

\newcommand{\ttenE}{\dot{\tenE}}
\newcommand{\ttenF}{\dot{\tenF}}
\newcommand{\ttenS}{\dot{\tenS}}
\newcommand{\ttenT}{\dot{\tenT}}
\newcommand{\ttenU}{\dot{\tenU}}
\newcommand{\ttenV}{\dot{\tenV}}
\newcommand{\ttenW}{\dot{\tenW}}
\newcommand{\ttenX}{\dot{\tenX}}
\newcommand{\ttenY}{\dot{\tenY}}
\newcommand{\ttenZ}{\dot{\tenZ}}

\newcommand{\ltenA}{\stackrel{\triangle}{\tenA}}
\newcommand{\ltenD}{\stackrel{\triangle}{\tenD}}
\newcommand{\ltenM}{\stackrel{\triangle}{\tenM}}

\newcommand{\btena}{\bar{\tena}}
\newcommand{\btenb}{\bar{\tenb}}
\newcommand{\btenc}{\bar{\tenc}}
\newcommand{\btend}{\bar{\tend}}
\newcommand{\btene}{\bar{\tene}}
\newcommand{\btenf}{\bar{\tenf}}
\newcommand{\bteng}{\bar{\teng}}

\newcommand{\btenA}{\bar{\tenA}}
\newcommand{\btenB}{\bar{\tenB}}
\newcommand{\btenC}{\bar{\tenC}}
\newcommand{\btenD}{\bar{\tenD}}
\newcommand{\btenE}{\bar{\tenE}}
\newcommand{\btenF}{\bar{\tenF}}
\newcommand{\btenK}{\bar{\tenK}}
\newcommand{\btenL}{\bar{\tenL}}
\newcommand{\btenM}{\bar{\tenM}}
\newcommand{\btenN}{\bar{\tenN}}
\newcommand{\btenO}{\bar{\tenO}}
\newcommand{\btenP}{\bar{\tenP}}
\newcommand{\btenQ}{\bar{\tenQ}}
\newcommand{\btenR}{\bar{\tenR}}
\newcommand{\btenS}{\bar{\tenS}}
\newcommand{\btenT}{\bar{\tenT}}


\newcommand{\lbtenE}{\Delta\bar{\tenE}}

\newcommand{\hattena}{\ensuremath{ \widehat{\ten{a}} }}
\newcommand{\hattenb}{\ensuremath{ \widehat{\ten{b}} }}
\newcommand{\hattenc}{\ensuremath{ \widehat{\ten{c}} }}
\newcommand{\hattend}{\ensuremath{ \widehat{\ten{d}} }}
\newcommand{\hattene}{\ensuremath{ \widehat{\ten{e}} }}
\newcommand{\hattenf}{\ensuremath{ \widehat{\ten{f}} }}
\newcommand{\hatteng}{\ensuremath{ \widehat{\ten{g}} }}
\newcommand{\hattenh}{\ensuremath{ \widehat{\ten{h}} }}
\newcommand{\hatteni}{\ensuremath{ \widehat{\ten{i}} }}
\newcommand{\hattenj}{\ensuremath{ \widehat{\ten{j}} }}
\newcommand{\hattenk}{\ensuremath{ \widehat{\ten{k}} }}
\newcommand{\hattenl}{\ensuremath{ \widehat{\ten{l}} }}
\newcommand{\hattenm}{\ensuremath{ \widehat{\ten{m}} }}
\newcommand{\hattenn}{\ensuremath{ \widehat{\ten{n}} }}
\newcommand{\hatteno}{\ensuremath{ \widehat{\ten{o}} }}
\newcommand{\hattenp}{\ensuremath{ \widehat{\ten{p}} }}
\newcommand{\hattenq}{\ensuremath{ \widehat{\ten{q}} }}
\newcommand{\hattenr}{\ensuremath{ \widehat{\ten{r}} }}
\newcommand{\hattens}{\ensuremath{ \widehat{\ten{s}} }}
\newcommand{\hattent}{\ensuremath{ \widehat{\ten{t}} }}
\newcommand{\hattenu}{\ensuremath{ \widehat{\ten{u}} }}
\newcommand{\hattenv}{\ensuremath{ \widehat{\ten{v}} }}
\newcommand{\hattenw}{\ensuremath{ \widehat{\ten{w}} }}
\newcommand{\hattenx}{\ensuremath{ \widehat{\ten{x}} }}
\newcommand{\hatteny}{\ensuremath{ \widehat{\ten{y}} }}
\newcommand{\hattenz}{\ensuremath{ \widehat{\ten{z}} }}

\newcommand{\hattenA}{\ensuremath{ \widehat{\ten{A}} }}
\newcommand{\hattenB}{\ensuremath{ \widehat{\ten{B}} }}
\newcommand{\hattenC}{\ensuremath{ \widehat{\ten{C}} }}
\newcommand{\hattenD}{\ensuremath{ \widehat{\ten{D}} }}
\newcommand{\hattenE}{\ensuremath{ \widehat{\ten{E}} }}
\newcommand{\hattenF}{\ensuremath{ \widehat{\ten{F}} }}
\newcommand{\hattenG}{\ensuremath{ \widehat{\ten{G}} }}
\newcommand{\hattenH}{\ensuremath{ \widehat{\ten{H}} }}
\newcommand{\hattenI}{\ensuremath{ \widehat{\ten{I}} }}
\newcommand{\hattenJ}{\ensuremath{ \widehat{\ten{J}} }}
\newcommand{\hattenK}{\ensuremath{ \widehat{\ten{K}} }}
\newcommand{\hattenL}{\ensuremath{ \widehat{\ten{L}} }}
\newcommand{\hattenM}{\ensuremath{ \widehat{\ten{M}} }}
\newcommand{\hattenN}{\ensuremath{ \widehat{\ten{N}} }}
\newcommand{\hattenO}{\ensuremath{ \widehat{\ten{O}} }}
\newcommand{\hattenP}{\ensuremath{ \widehat{\ten{P}} }}
\newcommand{\hattenQ}{\ensuremath{ \widehat{\ten{Q}} }}
\newcommand{\hattenR}{\ensuremath{ \widehat{\ten{R}} }}
\newcommand{\hattenS}{\ensuremath{ \widehat{\ten{S}} }}
\newcommand{\hattenT}{\ensuremath{ \widehat{\ten{T}} }}
\newcommand{\hattenU}{\ensuremath{ \widehat{\ten{U}} }}
\newcommand{\hattenV}{\ensuremath{ \widehat{\ten{V}} }}
\newcommand{\hattenW}{\ensuremath{ \widehat{\ten{W}} }}
\newcommand{\hattenX}{\ensuremath{ \widehat{\ten{X}} }}
\newcommand{\hattenY}{\ensuremath{ \widehat{\ten{Y}} }}
\newcommand{\hattenZ}{\ensuremath{ \widehat{\ten{Z}} }}

\newcommand{\tiltena}{\ensuremath{ \widetilde{\ten{a}} }}
\newcommand{\tiltenb}{\ensuremath{ \widetilde{\ten{b}} }}
\newcommand{\tiltenc}{\ensuremath{ \widetilde{\ten{c}} }}
\newcommand{\tiltend}{\ensuremath{ \widetilde{\ten{d}} }}
\newcommand{\tiltene}{\ensuremath{ \widetilde{\ten{e}} }}
\newcommand{\tiltenf}{\ensuremath{ \widetilde{\ten{f}} }}
\newcommand{\tilteng}{\ensuremath{ \widetilde{\ten{g}} }}
\newcommand{\tiltenh}{\ensuremath{ \widetilde{\ten{h}} }}
\newcommand{\tilteni}{\ensuremath{ \widetilde{\ten{i}} }}
\newcommand{\tiltenj}{\ensuremath{ \widetilde{\ten{j}} }}
\newcommand{\tiltenk}{\ensuremath{ \widetilde{\ten{k}} }}
\newcommand{\tiltenl}{\ensuremath{ \widetilde{\ten{l}} }}
\newcommand{\tiltenm}{\ensuremath{ \widetilde{\ten{m}} }}
\newcommand{\tiltenn}{\ensuremath{ \widetilde{\ten{n}} }}
\newcommand{\tilteno}{\ensuremath{ \widetilde{\ten{o}} }}
\newcommand{\tiltenp}{\ensuremath{ \widetilde{\ten{p}} }}
\newcommand{\tiltenq}{\ensuremath{ \widetilde{\ten{q}} }}
\newcommand{\tiltenr}{\ensuremath{ \widetilde{\ten{r}} }}
\newcommand{\tiltens}{\ensuremath{ \widetilde{\ten{s}} }}
\newcommand{\tiltent}{\ensuremath{ \widetilde{\ten{t}} }}
\newcommand{\tiltenu}{\ensuremath{ \widetilde{\ten{u}} }}
\newcommand{\tiltenv}{\ensuremath{ \widetilde{\ten{v}} }}
\newcommand{\tiltenw}{\ensuremath{ \widetilde{\ten{w}} }}
\newcommand{\tiltenx}{\ensuremath{ \widetilde{\ten{x}} }}
\newcommand{\tilteny}{\ensuremath{ \widetilde{\ten{y}} }}
\newcommand{\tiltenz}{\ensuremath{ \widetilde{\ten{z}} }}

\newcommand{\tiltenA}{\ensuremath{ \widetilde{\ten{A}} }}
\newcommand{\tiltenB}{\ensuremath{ \widetilde{\ten{B}} }}
\newcommand{\tiltenC}{\ensuremath{ \widetilde{\ten{C}} }}
\newcommand{\tiltenD}{\ensuremath{ \widetilde{\ten{D}} }}
\newcommand{\tiltenE}{\ensuremath{ \widetilde{\ten{E}} }}
\newcommand{\tiltenF}{\ensuremath{ \widetilde{\ten{F}} }}
\newcommand{\tiltenG}{\ensuremath{ \widetilde{\ten{G}} }}
\newcommand{\tiltenH}{\ensuremath{ \widetilde{\ten{H}} }}
\newcommand{\tiltenI}{\ensuremath{ \widetilde{\ten{I}} }}
\newcommand{\tiltenJ}{\ensuremath{ \widetilde{\ten{J}} }}
\newcommand{\tiltenK}{\ensuremath{ \widetilde{\ten{K}} }}
\newcommand{\tiltenL}{\ensuremath{ \widetilde{\ten{L}} }}
\newcommand{\tiltenM}{\ensuremath{ \widetilde{\ten{M}} }}
\newcommand{\tiltenN}{\ensuremath{ \widetilde{\ten{N}} }}
\newcommand{\tiltenO}{\ensuremath{ \widetilde{\ten{O}} }}
\newcommand{\tiltenP}{\ensuremath{ \widetilde{\ten{P}} }}
\newcommand{\tiltenQ}{\ensuremath{ \widetilde{\ten{Q}} }}
\newcommand{\tiltenR}{\ensuremath{ \widetilde{\ten{R}} }}
\newcommand{\tiltenS}{\ensuremath{ \widetilde{\ten{S}} }}
\newcommand{\tiltenT}{\ensuremath{ \widetilde{\ten{T}} }}
\newcommand{\tiltenU}{\ensuremath{ \widetilde{\ten{U}} }}
\newcommand{\tiltenV}{\ensuremath{ \widetilde{\ten{V}} }}
\newcommand{\tiltenW}{\ensuremath{ \widetilde{\ten{W}} }}
\newcommand{\tiltenX}{\ensuremath{ \widetilde{\ten{X}} }}
\newcommand{\tiltenY}{\ensuremath{ \widetilde{\ten{Y}} }}
\newcommand{\tiltenZ}{\ensuremath{ \widetilde{\ten{Z}} }}

\newcommand{\veca}{\ensuremath{ \vec{a} }}
\newcommand{\vecb}{\ensuremath{ \vec{b} }}
\newcommand{\vecc}{\ensuremath{ \vec{c} }}
\newcommand{\vecd}{\ensuremath{ \vec{d} }}
\newcommand{\vece}{\ensuremath{ \vec{e} }}
\newcommand{\vecf}{\ensuremath{ \vec{f} }}
\newcommand{\vecg}{\ensuremath{ \vec{g} }}
\newcommand{\vech}{\ensuremath{ \vec{h} }}
\newcommand{\veci}{\ensuremath{ \vec{i} }}
\newcommand{\vecj}{\ensuremath{ \vec{j} }}
\newcommand{\veck}{\ensuremath{ \vec{k} }}
\newcommand{\vecl}{\ensuremath{ \vec{l} }}
\newcommand{\vecm}{\ensuremath{ \vec{m} }}
\newcommand{\vecn}{\ensuremath{ \vec{n} }}
\newcommand{\veco}{\ensuremath{ \vec{o} }}
\newcommand{\vecp}{\ensuremath{ \vec{p} }}
\newcommand{\vecq}{\ensuremath{ \vec{q} }}
\newcommand{\vecr}{\ensuremath{ \vec{r} }}
\newcommand{\vecs}{\ensuremath{ \vec{s} }}
\newcommand{\vect}{\ensuremath{ \vec{t} }}
\newcommand{\vecu}{\ensuremath{ \vec{u} }}
\newcommand{\vecv}{\ensuremath{ \vec{v} }}
\newcommand{\vecw}{\ensuremath{ \vec{w} }}
\newcommand{\vecx}{\ensuremath{ \vec{x} }}
\newcommand{\vecy}{\ensuremath{ \vec{y} }}
\newcommand{\vecz}{\ensuremath{ \vec{z} }}

\newcommand{\vecA}{\ensuremath{ \vec{A} }}
\newcommand{\vecB}{\ensuremath{ \vec{B} }}
\newcommand{\vecC}{\ensuremath{ \vec{C} }}
\newcommand{\vecD}{\ensuremath{ \vec{D} }}
\newcommand{\vecE}{\ensuremath{ \vec{E} }}
\newcommand{\vecF}{\ensuremath{ \vec{F} }}
\newcommand{\vecG}{\ensuremath{ \vec{G} }}
\newcommand{\vecH}{\ensuremath{ \vec{H} }}
\newcommand{\vecI}{\ensuremath{ \vec{I} }}
\newcommand{\vecJ}{\ensuremath{ \vec{J} }}
\newcommand{\vecK}{\ensuremath{ \vec{K} }}
\newcommand{\vecL}{\ensuremath{ \vec{L} }}
\newcommand{\vecM}{\ensuremath{ \vec{M} }}
\newcommand{\vecN}{\ensuremath{ \vec{N} }}
\newcommand{\vecO}{\ensuremath{ \vec{O} }}
\newcommand{\vecP}{\ensuremath{ \vec{P} }}
\newcommand{\vecQ}{\ensuremath{ \vec{Q} }}
\newcommand{\vecR}{\ensuremath{ \vec{R} }}
\newcommand{\vecS}{\ensuremath{ \vec{S} }}
\newcommand{\vecT}{\ensuremath{ \vec{T} }}
\newcommand{\vecU}{\ensuremath{ \vec{U} }}
\newcommand{\vecV}{\ensuremath{ \vec{V} }}
\newcommand{\vecW}{\ensuremath{ \vec{W} }}
\newcommand{\vecX}{\ensuremath{ \vec{X} }}
\newcommand{\vecY}{\ensuremath{ \vec{Y} }}
\newcommand{\vecZ}{\ensuremath{ \vec{Z} }}

\newcommand{\tveca}{\ensuremath{ \dot{\vec{a}} }}
\newcommand{\tvecb}{\dot{\vecb}}
\newcommand{\tvecc}{\ensuremath{ \vec{c} }}
\newcommand{\tvecd}{\ensuremath{ \vec{d} }}
\newcommand{\tvece}{\ensuremath{ \vec{e} }}
\newcommand{\tvecf}{\ensuremath{ \vec{f} }}
\newcommand{\tvecg}{\ensuremath{ \dot{\vec{g}} }}
\newcommand{\tvech}{\ensuremath{ \vec{h} }}
\newcommand{\tveci}{\ensuremath{ \vec{i} }}
\newcommand{\tvecj}{\ensuremath{ \vec{j} }}
\newcommand{\tveck}{\ensuremath{ \vec{k} }}
\newcommand{\tvecl}{\ensuremath{ \vec{l} }}
\newcommand{\tvecm}{\ensuremath{ \vec{m} }}
\newcommand{\tvecn}{\dot{\ensuremath{ \vec{n} }}}
\newcommand{\tveco}{\dot{\ensuremath{ \vec{o} }}}
\newcommand{\tvecp}{\dot{\ensuremath{ \vec{p} }}}
\newcommand{\tvecq}{\dot{\ensuremath{ \vec{q} }}}
\newcommand{\tvecr}{\ensuremath{ \vec{r} }}
\newcommand{\tvecs}{\ensuremath{ \vec{s} }}
\newcommand{\tvect}{\ensuremath{ \vec{t} }}
\newcommand{\tvecu}{\dot{\ensuremath{ \vec{u} }}}
\newcommand{\tvecv}{\dot{\ensuremath{ \vec{v} }}}
\newcommand{\tvecw}{\dot{\ensuremath{ \vec{w} }}}
\newcommand{\tvecx}{\dot{\ensuremath{ \vec{x} }}}
\newcommand{\tvecy}{\dot{\ensuremath{ \vec{y} }}}
\newcommand{\tvecz}{\dot{\ensuremath{ \vec{z} }}}

\newcommand{\tvecA}{\ensuremath{ \vec{A} }}
\newcommand{\tvecB}{\ensuremath{ \vec{B} }}
\newcommand{\tvecC}{\ensuremath{ \vec{C} }}
\newcommand{\tvecD}{\ensuremath{ \vec{D} }}
\newcommand{\tvecE}{\ensuremath{ \vec{E} }}
\newcommand{\tvecF}{\ensuremath{ \vec{F} }}
\newcommand{\tvecG}{\ensuremath{ \vec{G} }}
\newcommand{\tvecH}{\ensuremath{ \vec{H} }}
\newcommand{\tvecI}{\ensuremath{ \vec{I} }}
\newcommand{\tvecJ}{\ensuremath{ \vec{J} }}
\newcommand{\tvecK}{\ensuremath{ \vec{K} }}
\newcommand{\tvecL}{\ensuremath{ \vec{L} }}
\newcommand{\tvecM}{\ensuremath{ \vec{M} }}
\newcommand{\tvecN}{\ensuremath{ \vec{N} }}
\newcommand{\tvecO}{\ensuremath{ \vec{O} }}
\newcommand{\tvecP}{\ensuremath{ \vec{P} }}
\newcommand{\tvecQ}{\ensuremath{ \vec{Q} }}
\newcommand{\tvecR}{\ensuremath{ \vec{R} }}
\newcommand{\tvecS}{\ensuremath{ \vec{S} }}
\newcommand{\tvecT}{\ensuremath{ \vec{T} }}
\newcommand{\tvecU}{\ensuremath{ \vec{U} }}
\newcommand{\tvecV}{\ensuremath{ \vec{V} }}
\newcommand{\tvecW}{\ensuremath{ \vec{W} }}
\newcommand{\tvecX}{\ensuremath{ \vec{X} }}
\newcommand{\tvecY}{\ensuremath{ \vec{Y} }}
\newcommand{\tvecZ}{\ensuremath{ \vec{Z} }}

\newcommand{\no}{\hspace{-0.14cm}\,}

\newcommand{\lveca}{\Delta\veca}
\newcommand{\lvecb}{\Delta\vecb}
\newcommand{\lvecc}{\Delta\vecc}
\newcommand{\lvecd}{\Delta\vecd}
\newcommand{\lvece}{\Delta\vece}
\newcommand{\lvecf}{\Delta\vecf}
\newcommand{\lvecg}{\Delta\vecg}
\newcommand{\lvech}{\Delta\vech}
\newcommand{\lveci}{\Delta\veci}
\newcommand{\lvecj}{\Delta\vecj}
\newcommand{\lveck}{\Delta\veck}
\newcommand{\lvecl}{\Delta\vecl}
\newcommand{\lvecm}{\Delta\vecm}
\newcommand{\lvecn}{\Delta\vecn}
\newcommand{\lveco}{\Delta\veco}
\newcommand{\lvecp}{\Delta\vecp}
\newcommand{\lvecq}{\Delta\vecq}
\newcommand{\lvecr}{\Delta\vecr}
\newcommand{\lvecs}{\Delta\vecs}
\newcommand{\lvect}{\Delta\vect}
\newcommand{\lvecu}{\Delta\vecu}
\newcommand{\lvecv}{\Delta\vecv}
\newcommand{\lvecw}{\Delta\vecw}
\newcommand{\lvecx}{\Delta\vecx}
\newcommand{\lvecy}{\Delta\vecy}
\newcommand{\lvecz}{\Delta\vecz}

\newcommand{\lvecN}{\Delta\vecN}


\newcommand{\dveca}{\delta\veca}
\newcommand{\dvecb}{\delta\vecb}
\newcommand{\dvecc}{\delta\vecc}
\newcommand{\dvecd}{\delta\vecd}
\newcommand{\dvece}{\delta\vece}
\newcommand{\dvecf}{\delta\vecf}
\newcommand{\dvecg}{\delta\vecg}
\newcommand{\dvech}{\delta\vech}
\newcommand{\dveci}{\delta\veci}
\newcommand{\dvecj}{\delta\vecj}
\newcommand{\dveck}{\delta\veck}
\newcommand{\dvecl}{\delta\vecl}
\newcommand{\dvecm}{\delta\vecm}
\newcommand{\dvecu}{\delta\vecu}
\newcommand{\dvecv}{\delta\vecv}
\newcommand{\dvecw}{\delta\vecw}
\newcommand{\dvecx}{\delta\vecx}
\newcommand{\dvecy}{\delta\vecy}
\newcommand{\dvecz}{\delta\vecz}

\newcommand{\bveca}{\bar{\veca}}
\newcommand{\bvecb}{\bar{\vecb}}
\newcommand{\bvecc}{\bar{\vecc}}
\newcommand{\bvecd}{\bar{\vecd}}
\newcommand{\bvece}{\bar{\vece}}
\newcommand{\bvecf}{\bar{\vecf}}
\newcommand{\bvecg}{\bar{\vecg}}
\newcommand{\bvecm}{\bar{\vecm}}
\newcommand{\bvecn}{\bar{\vecn}}
\newcommand{\bveco}{\bar{\veco}}
\newcommand{\bvecp}{\bar{\vecp}}
\newcommand{\bvecq}{\bar{\vecq}}
\newcommand{\bvecr}{\bar{\vecr}}
\newcommand{\bvecs}{\bar{\vecs}}
\newcommand{\bvect}{\bar{\vect}}
\newcommand{\bvecu}{\bar{\vecu}}
\newcommand{\bvecv}{\bar{\vecv}}
\newcommand{\bvecw}{\bar{\vecw}}
\newcommand{\bvecx}{\bar{\vecx}}
\newcommand{\bvecy}{\bar{\vecy}}
\newcommand{\bvecz}{\bar{\vecz}}

\newcommand{\bvecA}{\bar{\vecA}}
\newcommand{\bvecB}{\bar{\vecB}}
\newcommand{\bvecC}{\bar{\vecC}}
\newcommand{\bvecD}{\bar{\vecD}}
\newcommand{\bvecE}{\bar{\vecE}}
\newcommand{\bvecF}{\bar{\vecF}}
\newcommand{\bvecG}{\bar{\vecG}}
\newcommand{\bvecH}{\bar{\vecH}}
\newcommand{\bvecI}{\bar{\vecI}}
\newcommand{\bvecJ}{\bar{\vecJ}}
\newcommand{\bvecK}{\bar{\vecK}}
\newcommand{\bvecL}{\bar{\vecL}}
\newcommand{\bvecM}{\bar{\vecM}}
\newcommand{\bvecN}{\bar{\vecN}}
\newcommand{\bvecO}{\bar{\vecO}}
\newcommand{\bvecP}{\bar{\vecP}}
\newcommand{\bvecR}{\bar{\vecR}}
\newcommand{\bvecS}{\bar{\vecS}}
\newcommand{\bvecT}{\bar{\vecT}}
\newcommand{\bvecU}{\bar{\vecU}}


\newcommand{\lbveca}{\Delta\bar{\veca}}
\newcommand{\lbvecb}{\Delta\bar{\vecb}}
\newcommand{\lbvecc}{\Delta\bar{\vecc}}
\newcommand{\lbvecd}{\Delta\bar{\vecd}}
\newcommand{\lbvece}{\Delta\bar{\vece}}
\newcommand{\lbvecf}{\Delta\bar{\vecf}}
\newcommand{\lbvecg}{\Delta\bar{\vecg}}
\newcommand{\lbvech}{\Delta\bar{\vech}}
\newcommand{\lbveci}{\Delta\bar{\veci}}
\newcommand{\lbvecj}{\Delta\bar{\vecj}}
\newcommand{\lbvect}{\Delta\bar{\vect}}
\newcommand{\lbvecu}{\Delta\bar{\vecu}}
\newcommand{\lbvecx}{\Delta\bar{\vecx}}


\newcommand{\dbveca}{\delta\bar{\veca}}
\newcommand{\dbvecb}{\delta\bar{\vecb}}
\newcommand{\dbvecc}{\delta\bar{\vecc}}
\newcommand{\dbvecd}{\delta\bar{\vecd}}
\newcommand{\dbvece}{\delta\bar{\vece}}
\newcommand{\dbvecf}{\delta\bar{\vecf}}
\newcommand{\dbvecg}{\delta\bar{\vecg}}
\newcommand{\dbvech}{\delta\bar{\vech}}
\newcommand{\dbveci}{\delta\bar{\veci}}
\newcommand{\dbvecj}{\delta\bar{\vecj}}
\newcommand{\dbvect}{\delta\bar{\vect}}
\newcommand{\dbvecu}{\delta\bar{\vecu}}
\newcommand{\dbvecx}{\delta\bar{\vecx}}

\newcommand{\hatveca}{\ensuremath{ \widehat{\vec{a}} }}
\newcommand{\hatvecb}{\ensuremath{ \widehat{\vec{b}} }}
\newcommand{\hatvecc}{\ensuremath{ \widehat{\vec{c}} }}
\newcommand{\hatvecd}{\ensuremath{ \widehat{\vec{d}} }}
\newcommand{\hatvece}{\ensuremath{ \widehat{\vec{e}} }}
\newcommand{\hatvecf}{\ensuremath{ \widehat{\vec{f}} }}
\newcommand{\hatvecg}{\ensuremath{ \widehat{\vec{g}} }}
\newcommand{\hatvech}{\ensuremath{ \widehat{\vec{h}} }}
\newcommand{\hatveci}{\ensuremath{ \widehat{\vec{i}} }}
\newcommand{\hatvecj}{\ensuremath{ \widehat{\vec{j}} }}
\newcommand{\hatveck}{\ensuremath{ \widehat{\vec{k}} }}
\newcommand{\hatvecl}{\ensuremath{ \widehat{\vec{l}} }}
\newcommand{\hatvecm}{\ensuremath{ \widehat{\vec{m}} }}
\newcommand{\hatvecn}{\ensuremath{ \widehat{\vec{n}} }}
\newcommand{\hatveco}{\ensuremath{ \widehat{\vec{o}} }}
\newcommand{\hatvecp}{\ensuremath{ \widehat{\vec{p}} }}
\newcommand{\hatvecq}{\ensuremath{ \widehat{\vec{q}} }}
\newcommand{\hatvecr}{\ensuremath{ \widehat{\vec{r}} }}
\newcommand{\hatvecs}{\ensuremath{ \widehat{\vec{s}} }}
\newcommand{\hatvect}{\ensuremath{ \widehat{\vec{t}} }}
\newcommand{\hatvecu}{\ensuremath{ \widehat{\vec{u}} }}
\newcommand{\hatvecv}{\ensuremath{ \widehat{\vec{v}} }}
\newcommand{\hatvecw}{\ensuremath{ \widehat{\vec{w}} }}
\newcommand{\hatvecx}{\ensuremath{ \widehat{\vec{x}} }}
\newcommand{\hatvecy}{\ensuremath{ \widehat{\vec{y}} }}
\newcommand{\hatvecz}{\ensuremath{ \widehat{\vec{z}} }}

\newcommand{\hatvecA}{\ensuremath{ \widehat{\vec{A}} }}
\newcommand{\hatvecB}{\ensuremath{ \widehat{\vec{B}} }}
\newcommand{\hatvecC}{\ensuremath{ \widehat{\vec{C}} }}
\newcommand{\hatvecD}{\ensuremath{ \widehat{\vec{D}} }}
\newcommand{\hatvecE}{\ensuremath{ \widehat{\vec{E}} }}
\newcommand{\hatvecF}{\ensuremath{ \widehat{\vec{F}} }}
\newcommand{\hatvecG}{\ensuremath{ \widehat{\vec{G}} }}
\newcommand{\hatvecH}{\ensuremath{ \widehat{\vec{H}} }}
\newcommand{\hatvecI}{\ensuremath{ \widehat{\vec{I}} }}
\newcommand{\hatvecJ}{\ensuremath{ \widehat{\vec{J}} }}
\newcommand{\hatvecK}{\ensuremath{ \widehat{\vec{K}} }}
\newcommand{\hatvecL}{\ensuremath{ \widehat{\vec{L}} }}
\newcommand{\hatvecM}{\ensuremath{ \widehat{\vec{M}} }}
\newcommand{\hatvecN}{\ensuremath{ \widehat{\vec{N}} }}
\newcommand{\hatvecO}{\ensuremath{ \widehat{\vec{O}} }}
\newcommand{\hatvecP}{\ensuremath{ \widehat{\vec{P}} }}
\newcommand{\hatvecQ}{\ensuremath{ \widehat{\vec{Q}} }}
\newcommand{\hatvecR}{\ensuremath{ \widehat{\vec{R}} }}
\newcommand{\hatvecS}{\ensuremath{ \widehat{\vec{S}} }}
\newcommand{\hatvecT}{\ensuremath{ \widehat{\vec{T}} }}
\newcommand{\hatvecU}{\ensuremath{ \widehat{\vec{U}} }}
\newcommand{\hatvecV}{\ensuremath{ \widehat{\vec{V}} }}
\newcommand{\hatvecW}{\ensuremath{ \widehat{\vec{W}} }}
\newcommand{\hatvecX}{\ensuremath{ \widehat{\vec{X}} }}
\newcommand{\hatvecY}{\ensuremath{ \widehat{\vec{Y}} }}
\newcommand{\hatvecZ}{\ensuremath{ \widehat{\vec{Z}} }}

\newcommand{\tilveca}{\ensuremath{ \widetilde{\vec{a}} }}
\newcommand{\tilvecb}{\ensuremath{ \widetilde{\vec{b}} }}
\newcommand{\tilvecc}{\ensuremath{ \widetilde{\vec{c}} }}
\newcommand{\tilvecd}{\ensuremath{ \widetilde{\vec{d}} }}
\newcommand{\tilvece}{\ensuremath{ \widetilde{\vec{e}} }}
\newcommand{\tilvecf}{\ensuremath{ \widetilde{\vec{f}} }}
\newcommand{\tilvecg}{\ensuremath{ \widetilde{\vec{g}} }}
\newcommand{\tilvech}{\ensuremath{ \widetilde{\vec{h}} }}
\newcommand{\tilveci}{\ensuremath{ \widetilde{\vec{i}} }}
\newcommand{\tilvecj}{\ensuremath{ \widetilde{\vec{j}} }}
\newcommand{\tilveck}{\ensuremath{ \widetilde{\vec{k}} }}
\newcommand{\tilvecl}{\ensuremath{ \widetilde{\vec{l}} }}
\newcommand{\tilvecm}{\ensuremath{ \widetilde{\vec{m}} }}
\newcommand{\tilvecn}{\ensuremath{ \widetilde{\vec{n}} }}
\newcommand{\tilveco}{\ensuremath{ \widetilde{\vec{o}} }}
\newcommand{\tilvecp}{\ensuremath{ \widetilde{\vec{p}} }}
\newcommand{\tilvecq}{\ensuremath{ \widetilde{\vec{q}} }}
\newcommand{\tilvecr}{\ensuremath{ \widetilde{\vec{r}} }}
\newcommand{\tilvecs}{\ensuremath{ \widetilde{\vec{s}} }}
\newcommand{\tilvect}{\ensuremath{ \widetilde{\vec{t}} }}
\newcommand{\tilvecu}{\ensuremath{ \widetilde{\vec{u}} }}
\newcommand{\tilvecv}{\ensuremath{ \widetilde{\vec{v}} }}
\newcommand{\tilvecw}{\ensuremath{ \widetilde{\vec{w}} }}
\newcommand{\tilvecx}{\ensuremath{ \widetilde{\vec{x}} }}
\newcommand{\tilvecy}{\ensuremath{ \widetilde{\vec{y}} }}
\newcommand{\tilvecz}{\ensuremath{ \widetilde{\vec{z}} }}

\newcommand{\tilvecA}{\ensuremath{ \widetilde{\vec{A}} }}
\newcommand{\tilvecB}{\ensuremath{ \widetilde{\vec{B}} }}
\newcommand{\tilvecC}{\ensuremath{ \widetilde{\vec{C}} }}
\newcommand{\tilvecD}{\ensuremath{ \widetilde{\vec{D}} }}
\newcommand{\tilvecE}{\ensuremath{ \widetilde{\vec{E}} }}
\newcommand{\tilvecF}{\ensuremath{ \widetilde{\vec{F}} }}
\newcommand{\tilvecG}{\ensuremath{ \widetilde{\vec{G}} }}
\newcommand{\tilvecH}{\ensuremath{ \widetilde{\vec{H}} }}
\newcommand{\tilvecI}{\ensuremath{ \widetilde{\vec{I}} }}
\newcommand{\tilvecJ}{\ensuremath{ \widetilde{\vec{J}} }}
\newcommand{\tilvecK}{\ensuremath{ \widetilde{\vec{K}} }}
\newcommand{\tilvecL}{\ensuremath{ \widetilde{\vec{L}} }}
\newcommand{\tilvecM}{\ensuremath{ \widetilde{\vec{M}} }}
\newcommand{\tilvecN}{\ensuremath{ \widetilde{\vec{N}} }}
\newcommand{\tilvecO}{\ensuremath{ \widetilde{\vec{O}} }}
\newcommand{\tilvecP}{\ensuremath{ \widetilde{\vec{P}} }}
\newcommand{\tilvecQ}{\ensuremath{ \widetilde{\vec{Q}} }}
\newcommand{\tilvecR}{\ensuremath{ \widetilde{\vec{R}} }}
\newcommand{\tilvecS}{\ensuremath{ \widetilde{\vec{S}} }}
\newcommand{\tilvecT}{\ensuremath{ \widetilde{\vec{T}} }}
\newcommand{\tilvecU}{\ensuremath{ \widetilde{\vec{U}} }}
\newcommand{\tilvecV}{\ensuremath{ \widetilde{\vec{V}} }}
\newcommand{\tilvecW}{\ensuremath{ \widetilde{\vec{W}} }}
\newcommand{\tilvecX}{\ensuremath{ \widetilde{\vec{X}} }}
\newcommand{\tilvecY}{\ensuremath{ \widetilde{\vec{Y}} }}
\newcommand{\tilvecZ}{\ensuremath{ \widetilde{\vec{Z}} }}

\newcommand{\tilveclam}{\ensuremath{ \widetilde{\gvec{\lambda}} }}

\newcommand{\ttvecd}{\ensuremath{ \ddot{\vec{d}} }}
\newcommand{\ttvect}{\ensuremath{ \ddot{\vec{t}} }}
\newcommand{\ttvecu}{\ensuremath{ \ddot{\vec{u}} }}
\newcommand{\ttvecv}{\ensuremath{ \ddot{\vec{v}} }}
\newcommand{\ttvecw}{\ensuremath{ \ddot{\vec{w}} }}
\newcommand{\ttvecx}{\ensuremath{ \ddot{\vec{x}} }}
\newcommand{\ttvecy}{\ensuremath{ \ddot{\vec{y}} }}
\newcommand{\ttvecz}{\ensuremath{ \ddot{\vec{z}} }}

\newcommand{\scaa}{\ensuremath{\mathrm{a}}}
\newcommand{\scab}{\ensuremath{\mathrm{b}}}
\newcommand{\scac}{\ensuremath{\mathrm{c}}}
\newcommand{\scad}{\ensuremath{\mathrm{d}}}
\newcommand{\scae}{\ensuremath{\mathrm{e}}}
\newcommand{\scaf}{\ensuremath{\mathrm{f}}}
\newcommand{\scag}{\ensuremath{\mathrm{g}}}
\newcommand{\scah}{\ensuremath{\mathrm{h}}}
\newcommand{\scai}{\ensuremath{\mathrm{i}}}
\newcommand{\scaj}{\ensuremath{\mathrm{j}}}
\newcommand{\scak}{\ensuremath{\mathrm{k}}}
\newcommand{\scal}{\ensuremath{\mathrm{l}}}
\newcommand{\scam}{\ensuremath{\mathrm{m}}}
\newcommand{\scan}{\ensuremath{\mathrm{n}}}
\newcommand{\scao}{\ensuremath{\mathrm{o}}}
\newcommand{\scap}{\ensuremath{\mathrm{p}}}
\newcommand{\scaq}{\ensuremath{\mathrm{q}}}
\newcommand{\scar}{\ensuremath{\mathrm{r}}}
\newcommand{\scas}{\ensuremath{\mathrm{s}}}
\newcommand{\scat}{\ensuremath{\mathrm{t}}}
\newcommand{\scau}{\ensuremath{\mathrm{u}}}
\newcommand{\scav}{\ensuremath{\mathrm{v}}}
\newcommand{\scaw}{\ensuremath{\mathrm{w}}}
\newcommand{\scax}{\ensuremath{\mathrm{x}}}
\newcommand{\scay}{\ensuremath{\mathrm{y}}}
\newcommand{\scaz}{\ensuremath{\mathrm{z}}}

\newcommand{\scaA}{\ensuremath{\mathrm{A}}}
\newcommand{\scaB}{\ensuremath{\mathrm{B}}}
\newcommand{\scaC}{\ensuremath{\mathrm{C}}}
\newcommand{\scaD}{\ensuremath{\mathrm{D}}}
\newcommand{\scaE}{\ensuremath{\mathrm{E}}}
\newcommand{\scaF}{\ensuremath{\mathrm{F}}}
\newcommand{\scaG}{\ensuremath{\mathrm{G}}}
\newcommand{\scaH}{\ensuremath{\mathrm{H}}}
\newcommand{\scaI}{\ensuremath{\mathrm{I}}}
\newcommand{\scaJ}{\ensuremath{\mathrm{J}}}
\newcommand{\scaK}{\ensuremath{\mathrm{K}}}
\newcommand{\scaL}{\ensuremath{\mathrm{L}}}
\newcommand{\scaM}{\ensuremath{\mathrm{M}}}
\newcommand{\scaN}{\ensuremath{\mathrm{N}}}
\newcommand{\scaO}{\ensuremath{\mathrm{O}}}
\newcommand{\scaP}{\ensuremath{\mathrm{P}}}
\newcommand{\scaQ}{\ensuremath{\mathrm{Q}}}
\newcommand{\scaR}{\ensuremath{\mathrm{R}}}
\newcommand{\scaS}{\ensuremath{\mathrm{S}}}
\newcommand{\scaT}{\ensuremath{\mathrm{T}}}
\newcommand{\scaU}{\ensuremath{\mathrm{U}}}
\newcommand{\scaV}{\ensuremath{\mathrm{V}}}
\newcommand{\scaW}{\ensuremath{\mathrm{W}}}
\newcommand{\scaX}{\ensuremath{\mathrm{X}}}
\newcommand{\scaY}{\ensuremath{\mathrm{Y}}}
\newcommand{\scaZ}{\ensuremath{\mathrm{Z}}}

\newcommand{\bscaa}{\bar{\scaa}}
\newcommand{\bscab}{\bar{\scab}}
\newcommand{\bscac}{\bar{\scac}}
\newcommand{\bscad}{\bar{\scad}}
\newcommand{\bscae}{\bar{\scae}}
\newcommand{\bscaf}{\bar{\scaf}}
\newcommand{\bscag}{\bar{\scag}}
\newcommand{\bscah}{\bar{\scah}}
\newcommand{\bscai}{\bar{\scai}}
\newcommand{\bscaj}{\bar{\scaj}}
\newcommand{\bscak}{\bar{\scak}}
\newcommand{\bscal}{\bar{\scal}}
\newcommand{\bscam}{\bar{\scam}}
\newcommand{\bscan}{\bar{\scan}}
\newcommand{\bscao}{\bar{\scao}}
\newcommand{\bscap}{\bar{\scap}}
\newcommand{\bscaq}{\bar{\scaq}}
\newcommand{\bscar}{\bar{\scar}}
\newcommand{\bscas}{\bar{\scas}}
\newcommand{\bscat}{\bar{\scat}}

\newcommand{\bscaA}{\bar{\scaA}}
\newcommand{\bscaB}{\bar{\scaB}}
\newcommand{\bscaC}{\bar{\scaC}}
\newcommand{\bscaD}{\bar{\scaD}}
\newcommand{\bscaE}{\bar{\scaE}}
\newcommand{\bscaF}{\bar{\scaF}}
\newcommand{\bscaG}{\bar{\scaG}}
\newcommand{\bscaH}{\bar{\scaH}}
\newcommand{\bscaI}{\bar{\scaI}}
\newcommand{\bscaJ}{\bar{\scaJ}}
\newcommand{\bscaK}{\bar{\scaK}}
\newcommand{\bscaL}{\bar{\scaL}}
\newcommand{\bscaM}{\bar{\scaM}}
\newcommand{\bscaN}{\bar{\scaN}}
\newcommand{\bscaO}{\bar{\scaO}}
\newcommand{\bscaP}{\bar{\scaP}}
\newcommand{\bscaQ}{\bar{\scaQ}}
\newcommand{\bscaR}{\bar{\scaR}}
\newcommand{\bscaS}{\bar{\scaS}}
\newcommand{\bscaT}{\bar{\scaT}}

\newcommand{\hatscan}{\ensuremath{\widehat{\scan}}}

\newcommand{\tilscag}{\ensuremath{\widetilde{\scag}}}
\newcommand{\tilscat}{\ensuremath{\widetilde{\scat}}}

\newcommand{\tillam}{\ensuremath{\widetilde{\lambda}}}

\newcommand{\tscag}{\ensuremath{\dot{\scag}}}

\newcommand{\lscaa}{\Delta\scaa}
\newcommand{\lscab}{\Delta\scab}
\newcommand{\lscac}{\Delta\scac}
\newcommand{\lscad}{\Delta\scad}
\newcommand{\lscae}{\Delta\scae}
\newcommand{\lscaf}{\Delta\scaf}
\newcommand{\lscag}{\Delta\scag}
\newcommand{\lscah}{\Delta\scah}
\newcommand{\lscai}{\Delta\scai}
\newcommand{\lscaj}{\Delta\scaj}
\newcommand{\lscak}{\Delta\scak}
\newcommand{\lscal}{\Delta\scal}
\newcommand{\lscam}{\Delta\scam}
\newcommand{\lscan}{\Delta\scan}
\newcommand{\lscao}{\Delta\scao}
\newcommand{\lscap}{\Delta\scap}
\newcommand{\lscaq}{\Delta\scaq}
\newcommand{\lscar}{\Delta\scar}
\newcommand{\lscas}{\Delta\scas}
\newcommand{\lscat}{\Delta\scat}

\newcommand{\lscaA}{\Delta\scaA}
\newcommand{\lscaD}{\Delta\scaD}
\newcommand{\lscaM}{\Delta\scaM}
\newcommand{\lscaN}{\Delta\scaN}

\newcommand{\balpha     }{\bar{\alpha}}
\newcommand{\bbeta      }{\bar{\beta}}
\newcommand{\bgamma     }{\bar{\gamma}}
\newcommand{\bdelta     }{\bar{\delta}}
\newcommand{\bepsilon   }{\bar{\epsilon}}
\newcommand{\bvareps    }{\bar{\varepsilon}}
\newcommand{\blambda    }{\bar{\lambda}}
\newcommand{\bxi        }{\bar{\xi}}
\newcommand{\bsigma     }{\bar{\sigma}}
\newcommand{\bvarsigma  }{\bar{\varsigma}}
\newcommand{\btau       }{\bar{\tau}}

\newcommand{\tileps     }{\widetilde{\epsilon}}
\newcommand{\tillambda  }{\widetilde{\lambda}}
\newcommand{\tilsigma   }{\widetilde{\sigma}}

\newcommand{\vecalpha     }{\ensuremath{ \gvec{\alpha} }}
\newcommand{\vecbeta      }{\ensuremath{ \gvec{\beta} }}
\newcommand{\vecgamma     }{\ensuremath{ \gvec{\gamma} }}
\newcommand{\vecdelta     }{\ensuremath{ \gvec{\delta} }}
\newcommand{\vecepsilon   }{\ensuremath{ \gvec{\epsilon} }}
\newcommand{\vecvarepsilon}{\ensuremath{ \gvec{\varepsilon} }}
\newcommand{\veczeta      }{\ensuremath{ \gvec{\zeta} }}
\newcommand{\veceta       }{\ensuremath{ \gvec{\eta} }}
\newcommand{\vectheta     }{\ensuremath{ \gvec{\theta} }}
\newcommand{\vecvartheta  }{\ensuremath{ \gvec{\vartheta} }}
\newcommand{\veciota      }{\ensuremath{ \gvec{\iota} }}
\newcommand{\veckappa     }{\ensuremath{ \gvec{\kappa} }}
\newcommand{\veclam       }{\ensuremath{ \gvec{\lambda} }}
\newcommand{\vecmu        }{\ensuremath{ \gvec{\mu} }}
\newcommand{\vecnu        }{\ensuremath{ \gvec{\nu} }}
\newcommand{\vecxi        }{\ensuremath{ \gvec{\xi} }}
\newcommand{\vecpi        }{\ensuremath{ \gvec{\pi} }}
\newcommand{\vecvarpi     }{\ensuremath{ \gvec{\varphi} }}
\newcommand{\vecrho       }{\ensuremath{ \gvec{\rho} }}
\newcommand{\vecvarrho    }{\ensuremath{ \gvec{\varrho} }}
\newcommand{\vecsigma     }{\ensuremath{ \gvec{\sigma} }}
\newcommand{\vecvarsigma  }{\ensuremath{ \gvec{\varsigma} }}
\newcommand{\vectau       }{\ensuremath{ \gvec{\tau} }}
\newcommand{\vecupsilon   }{\ensuremath{ \gvec{\upsilon} }}
\newcommand{\vecphi       }{\ensuremath{ \gvec{\phi} }}
\newcommand{\vecvarphi    }{\ensuremath{ \gvec{\varphi} }}
\newcommand{\vecchi       }{\ensuremath{ \gvec{\chi} }}
\newcommand{\vecpsi       }{\ensuremath{ \gvec{\psi} }}
\newcommand{\vecomega     }{\ensuremath{ \gvec{\omega} }}
\newcommand{\vecUpsilon   }{\ensuremath{ \gvec{\Upsilon} }}

\newcommand{\bveceps      }{\ensuremath{ \bar{\gvec{\epsilon}} }}
\newcommand{\bveceta      }{\ensuremath{ \bar{\gvec{\eta}} }}
\newcommand{\bveclam      }{\ensuremath{ \bar{\gvec{\lambda}} }}
\newcommand{\bvecsig      }{\ensuremath{ \bar{\gvec{\sigma}} }}
\newcommand{\bvecvarsigma }{\ensuremath{ \bar{\gvec{\varsigma}} }}
\newcommand{\bvectau      }{\ensuremath{ \bar{\gvec{\tau}} }}
\newcommand{\bvecupsilon  }{\ensuremath{ \bar{\gvec{\upsilon} }}}

\newcommand{\tilveceps    }{\widetilde{\vecepsilon}}
\newcommand{\tilvecsig    }{\widetilde{\vecsigma}}

\newcommand{\tveclam}{\ensuremath{ \dot{\veclam }}}

\newcommand{\lveceps}{\Delta\vecepsilon}
\newcommand{\lveclam}{\Delta\veclam}
\newcommand{\lvecsig}{\Delta\vecsigma}
\newcommand{\lvectau}{\Delta\vectau}
\newcommand{\lvecxi }{\Delta\vecxi}

\newcommand{\dveceps}{\delta\vecepsilon}
\newcommand{\dveclam}{\delta\veclam}
\newcommand{\dvecsig}{\delta\vecsigma}
\newcommand{\dvectau}{\delta\vectau}
\newcommand{\dvecxi }{\delta\vecxi}


\newcommand{\lbveclam}{\Delta\bar{\veclam}}

\newcommand{\tenalpha     }{\ensuremath{ \gten{\alpha} }}
\newcommand{\tenbeta      }{\ensuremath{ \gten{\beta} }}
\newcommand{\tengamma     }{\ensuremath{ \gten{\gamma} }}
\newcommand{\tendelta     }{\ensuremath{ \gten{\delta} }}
\newcommand{\tenepsilon   }{\ensuremath{ \gten{\epsilon} }}
\newcommand{\teneps       }{\ensuremath{ \gten{\varepsilon} }}
\newcommand{\tenzeta      }{\ensuremath{ \gten{\zeta} }}
\newcommand{\teneta       }{\ensuremath{ \gten{\eta} }}
\newcommand{\tentheta     }{\ensuremath{ \gten{\theta} }}
\newcommand{\tenvartheta  }{\ensuremath{ \gten{\vartheta} }}
\newcommand{\teniota      }{\ensuremath{ \gten{\iota} }}
\newcommand{\tenkappa     }{\ensuremath{ \gten{\kappa} }}
\newcommand{\tenlambda    }{\ensuremath{ \gten{\lambda} }}
\newcommand{\tenmu        }{\ensuremath{ \gten{\mu} }}
\newcommand{\tennu        }{\ensuremath{ \gten{\nu} }}
\newcommand{\tenxi        }{\ensuremath{ \gten{\xi} }}
\newcommand{\tenpi        }{\ensuremath{ \gten{\pi} }}
\newcommand{\tenvarpi     }{\ensuremath{ \gten{\varphi} }}
\newcommand{\tenrho       }{\ensuremath{ \gten{\rho} }}
\newcommand{\tenvarrho    }{\ensuremath{ \gten{\varrho} }}
\newcommand{\tensig       }{\ensuremath{ \gten{\sigma} }}
\newcommand{\tenvarsigma  }{\ensuremath{ \gten{\varsigma} }}
\newcommand{\tentau       }{\ensuremath{ \gten{\tau} }}
\newcommand{\tenupsilon   }{\ensuremath{ \gten{\upsilon} }}
\newcommand{\tenphi       }{\ensuremath{ \gten{\phi} }}
\newcommand{\tenvarphi    }{\ensuremath{ \gten{\varphi} }}
\newcommand{\tenchi       }{\ensuremath{ \gten{\chi} }}
\newcommand{\tenpsi       }{\ensuremath{ \gten{\psi} }}
\newcommand{\tenomega     }{\ensuremath{ \gten{\omega} }}

\newcommand{\tenOmega     }{\ensuremath{ \gten{\Omega} }}

\newcommand{\tilteneps    }{\widetilde{\teneps}}
\newcommand{\tiltensig    }{\widetilde{\tensig}}


\newcommand{\bteneps}{\ensuremath{ \bar{\teneps }}}
\newcommand{\btensig}{\ensuremath{ \bar{\tensig }}}


\newcommand{\tteneps}{\ensuremath{ \dot{\teneps }}}
\newcommand{\ttensig}{\ensuremath{ \dot{\tensig }}}


\newcommand{\ltenalpha}{\Delta\tenalpha}
\newcommand{\ltenbeta }{\Delta\tenbeta}
\newcommand{\lteneps  }{\Delta\teneps}
\newcommand{\ltensig  }{\Delta\tensig}

\newcommand{\tgamma}{\ensuremath{ \dot{\gamma} }}
\newcommand{\txi}{\ensuremath{ \dot{\xi} }}
\newcommand{\tlam}{\ensuremath{ \dot{\lambda} }}
\newcommand{\tomega}{\ensuremath{ \dot{\omega} }}

\newcommand{\lgamma}{\Delta\gamma}
\newcommand{\llambda}{\Delta\lambda}
\newcommand{\lxi}{\Delta\xi}
\newcommand{\lsigma}{\stackrel{\triangle}{\sigma}}
\newcommand{\ltau}{\stackrel{\triangle}{\tau}}

\newcommand{\hatalpha     }{\ensuremath{ \widehat{\alpha} }}
\newcommand{\hatbeta      }{\ensuremath{ \widehat{\beta} }}
\newcommand{\hatgamma     }{\ensuremath{ \widehat{\gamma} }}
\newcommand{\hatdelta     }{\ensuremath{ \widehat{\delta} }}
\newcommand{\hatepsilon   }{\ensuremath{ \widehat{\epsilon} }}
\newcommand{\hatvarepsilon}{\ensuremath{ \widehat{\varepsilon} }}
\newcommand{\hatzeta      }{\ensuremath{ \widehat{\zeta} }}
\newcommand{\hateta       }{\ensuremath{ \widehat{\eta} }}
\newcommand{\hattheta     }{\ensuremath{ \widehat{\theta} }}
\newcommand{\hatvartheta  }{\ensuremath{ \widehat{\vartheta} }}
\newcommand{\hatiota      }{\ensuremath{ \widehat{\iota} }}
\newcommand{\hatkappa     }{\ensuremath{ \widehat{\kappa} }}
\newcommand{\hatlambda    }{\ensuremath{ \widehat{\lambda} }}
\newcommand{\hatmu        }{\ensuremath{ \widehat{\mu} }}
\newcommand{\hatnu        }{\ensuremath{ \widehat{\nu} }}
\newcommand{\hatxi        }{\ensuremath{ \widehat{\xi} }}
\newcommand{\hatpi        }{\ensuremath{ \widehat{\pi} }}
\newcommand{\hatvarpi     }{\ensuremath{ \widehat{\varphi} }}
\newcommand{\hatrho       }{\ensuremath{ \widehat{\rho} }}
\newcommand{\hatvarrho    }{\ensuremath{ \widehat{\varrho} }}
\newcommand{\hatsigma     }{\ensuremath{ \widehat{\sigma} }}
\newcommand{\hatvarsigma  }{\ensuremath{ \widehat{\varsigma} }}
\newcommand{\hattau       }{\ensuremath{ \widehat{\tau} }}
\newcommand{\hatupsilon   }{\ensuremath{ \widehat{\upsilon} }}
\newcommand{\hatphi       }{\ensuremath{ \widehat{\phi} }}
\newcommand{\hatvarphi    }{\ensuremath{ \widehat{\varphi} }}
\newcommand{\hatchi       }{\ensuremath{ \widehat{\chi} }}
\newcommand{\hatpsi       }{\ensuremath{ \widehat{\psi} }}
\newcommand{\hatomega     }{\ensuremath{ \widehat{\omega} }}

\newcommand{\hatteneps}{\ensuremath{ \widehat{\teneps} }}
\newcommand{\hattensig}{\ensuremath{ \widehat{\tensig} }}

\newcommand{\ionesi}{\scaI\utensig}
\newcommand{\itwosi}{\scaI\scaI\utensig}
\newcommand{\ithrsi}{\scaI\scaI\scaI\utensig}
\newcommand{\itwos}{\scaI\scaI\utens}
\newcommand{\ithrs}{\scaI\scaI\scaI\utens}

\newcommand{\onetwo}{\frac{1}{2}}
\newcommand{\thrtwo}{\frac{3}{2}}
\newcommand{\onethr}{\frac{1}{3}}
\newcommand{\twothr}{\frac{2}{3}}
\newcommand{\forthr}{\frac{4}{3}}
\newcommand{\onefor}{\frac{1}{4}}
\newcommand{\onesix}{\frac{1}{6}}
\newcommand{\oneeig}{\frac{1}{8}}
\newcommand{\onenin}{\frac{1}{9}}
\newcommand{\onetwe}{\frac{1}{12}}

\newcommand{\tengf}{\teng^{\flat}}
\newcommand{\tengs}{\teng^{\sharp}}

\newcommand{\Lin}{^{Lin}}

\newcommand{\uscan}{_{\mbox{\tiny{N}}}}

\newcommand{\ena}{\ensuremath{^{n+1}}}
\newcommand{\sena}{\ensuremath{^{1\,n+1}}}
\newcommand{\mena}{\ensuremath{^{2\,n+1}}}

\newcommand{\ea}{^{\alpha}}
\newcommand{\eb}{^{\beta}}
\newcommand{\ec}{^{\gamma}}
\newcommand{\ed}{^{\delta}}
\newcommand{\ex}{^{\xi}}

\newcommand{\eat}{^{\alpha T}}
\newcommand{\ebt}{^{\beta T}}
\newcommand{\ect}{^{\gamma T}}
\newcommand{\edt}{^{\delta T}}
\newcommand{\eet}{^{\epsilon T}}

\newcommand{\eaa}{^{\alpha\alpha}}
\newcommand{\eab}{^{\alpha\beta}}
\newcommand{\eac}{^{\alpha\gamma}}
\newcommand{\ead}{^{\alpha\delta}}
\newcommand{\eba}{^{\beta\alpha}}
\newcommand{\ebc}{^{\beta\gamma}}
\newcommand{\ebd}{^{\beta\delta}}
\newcommand{\ecb}{^{\gamma\beta}}
\newcommand{\ecd}{^{\gamma\delta}}
\newcommand{\edb}{^{\delta\beta}}
\newcommand{\ede}{^{\delta\epsilon}}
\newcommand{\eec}{^{\epsilon\gamma}}
\newcommand{\exx}{^{\xi\xi}}

\newcommand{\ua}{_{\alpha}}
\newcommand{\ub}{_{\beta}}
\newcommand{\uc}{_{\gamma}}
\newcommand{\ud}{_{\delta}}
\newcommand{\ue}{_{\epsilon}}
\newcommand{\ux}{_{\xi}}

\newcommand{\uaa}{_{\alpha\alpha}}
\newcommand{\uab}{_{\alpha\beta}}
\newcommand{\uac}{_{\alpha\gamma}}
\newcommand{\uba}{_{\beta\alpha}}
\newcommand{\ubb}{_{\beta\beta}}
\newcommand{\ubc}{_{\beta\gamma}}
\newcommand{\ubd}{_{\beta\delta}}
\newcommand{\ucd}{_{\gamma\delta}}
\newcommand{\ucb}{_{\gamma\beta}}
\newcommand{\ueb}{_{\epsilon\beta}}
\newcommand{\ued}{_{\epsilon\delta}}
\newcommand{\udb}{_{\delta\beta}}
\newcommand{\uta}{_{{\mbox{\tiny{T}}}\alpha}}
\newcommand{\utb}{_{{\mbox{\tiny{T}}}\beta}}
\newcommand{\utc}{_{{\mbox{\tiny{T}}}\gamma}}
\newcommand{\uxx}{_{\xi\xi}}

\newcommand{\uka}{_{,\alpha}}
\newcommand{\ukb}{_{,\beta}}
\newcommand{\ukc}{_{,\gamma}}
\newcommand{\ukx}{_{,\xi}}

\newcommand{\uakb}{_{\alpha,\beta}}
\newcommand{\uakc}{_{\alpha,\gamma}}
\newcommand{\ubkc}{_{\beta,\gamma}}
\newcommand{\ubkd}{_{\beta,\delta}}
\newcommand{\ubke}{_{\beta,\epsilon}}
\newcommand{\uckd}{_{\gamma,\delta}}
\newcommand{\udke}{_{\delta,\epsilon}}

\newcommand{\ukaa}{_{,\alpha\alpha}}
\newcommand{\ukab}{_{,\alpha\beta}}
\newcommand{\ukba}{_{,\beta\alpha}}
\newcommand{\ukbb}{_{,\beta\beta}}
\newcommand{\ukbc}{_{,\beta\gamma}}
\newcommand{\ukxx}{_{,\xi\xi}}
\newcommand{\uxkx}{_{\xi,\xi}}
\newcommand{\uxkxx}{_{\xi,\xi\xi}}

\newcommand{\ubkcd}{_{\beta,\gamma\delta}}

\newcommand{\uga}{_{g\alpha}}
\newcommand{\ugb}{_{g\beta}}
\newcommand{\ugc}{_{g\gamma}}
\newcommand{\ugd}{_{g\delta}}
\newcommand{\ugka}{_{g,\alpha}}
\newcommand{\ugkb}{_{g,\beta}}
\newcommand{\ugkc}{_{g,\gamma}}
\newcommand{\ugkx}{_{g,\xi}}
\newcommand{\ugakb}{_{g\alpha,\beta}}
\newcommand{\ugbkc}{_{g\beta,\gamma}}

\newcommand{\uana}{_{\alpha\,n+1}}
\newcommand{\ubna}{_{\beta\,n+1}}

\newcommand{\ukana}{_{,\alpha\,n+1}}
\newcommand{\ukano}{_{,\alpha\,n}}

\newcommand{\uano}{_{\alpha\,n}}
\newcommand{\ubno}{_{\beta\,n}}

\newcommand{\umN}{_{\mbox{\tiny{N}}}}
\newcommand{\umT}{_{\mbox{\tiny{T}}}}

\newcommand{\utenb}{_{\tenb}}
\newcommand{\utenp}{_{\tenp}}
\newcommand{\utens}{_{\tens}}
\newcommand{\utenC}{_{\tenC}}
\newcommand{\utenE}{_{\tenE}}

\newcommand{\utensna}{_{\tens\,n+1}}

\newcommand{\uteneps}{_{\teneps}}
\newcommand{\utenepse}{_{\teneps^e}}
\newcommand{\utenepsp}{_{\teneps^p}}
\newcommand{\utensig}{_{\tensig}}
\newcommand{\utensigsig}{_{\tensig\tensig}}

\newcommand{\utenepsna}{_{\teneps\,n+1}}
\newcommand{\utensigna}{_{\tensig\,n+1}}

\newcommand{\gvecx}{\grave{\vecx}}


\newcommand{\mska}{_{s,\alpha}}
\newcommand{\mskb}{_{s,\beta}}
\newcommand{\mskc}{_{s,\gamma}}
\newcommand{\mfka}{_{f,\alpha}}
\newcommand{\mfkb}{_{f,\beta}}
\newcommand{\mfkc}{_{f,\gamma}}


\newcommand{\dmska}{_{Ag,\alpha}}
\newcommand{\dmskb}{_{Ag,\beta}}
\newcommand{\dmskc}{_{Ag,\gamma}}
\newcommand{\dmfka}{_{Ag,\alpha}}
\newcommand{\dmfkb}{_{Ag,\beta}}
\newcommand{\dmfkc}{_{Ag,\gamma}}

\newcommand{\gmska}{_{g,\alpha}}
\newcommand{\gmskb}{_{g,\beta}}
\newcommand{\gmskc}{_{g,\gamma}}
\newcommand{\gmfka}{_{g,\alpha}}
\newcommand{\gmfkb}{_{g,\beta}}
\newcommand{\gmfkc}{_{g,\gamma}}

\newcommand{\sumgp}{\sum_{g=1}^{n_{gp}}}
\newcommand{\sumni}{\sum_{I=1}^{n_{I}}}
\newcommand{\sumnj}{\sum_{J=1}^{n_{J}}}
\newcommand{\sumseg}{\sum_{seg}}
\newcommand{\sumel}{\sum_{e=1}^{n_{el}}}

\newcommand{\suma}{\sum_{\alpha=1}^{n_{\alpha}}}

\newcommand{\gng}{\scag_{N\,g}}
\newcommand{\gtg}{g_{T\,g}}
\newcommand{\lng}{\lambda_{N\,g}}
\newcommand{\ltg}{\lambda_{T\,g}}
\newcommand{\ttg}{t_{T\,g}}

\renewcommand{\d}[1]{\text{$\hspace{0.1cm}$d $\hspace{-0.11cm}#1$}}
\newcommand{\del}{\ensuremath{\partial}}
\newcommand{\divx}[1]{\text{$\hspace{0.1cm}$div$\left(#1\right)$}}
\newcommand{\divX}[1]{\text{$\hspace{0.1cm}$Div$\left(#1\right)$}}
\newcommand{\grad}[1]{\ensuremath{ \boldsymbol{\nabla}{#1}}}
\newcommand{\gradx}[1]{\ensuremath{ \boldsymbol{\nabla}_{\vecx}{#1}}}
\newcommand{\gradX}[1]{\ensuremath{ \boldsymbol{\nabla}_{\vecX}{#1}}}
\newcommand{\parder}[2]{\ensuremath{ \frac{\del #1}{\del #2} }}
\newcommand{\tder}[1]{\ensuremath{ \frac{\d{#1}}{\d{} \hspace{0.05cm}{t}} }}
\newcommand{\dx}{\ensuremath{ \d{\vecx} }}
\newcommand{\dX}{\ensuremath{ \d{\vecX} }}
\newcommand{\da}{\ensuremath{ \d{a} }}
\newcommand{\dA}{\ensuremath{ \d{A} }}
\newcommand{\dv}{\ensuremath{ \d{v} }}
\newcommand{\dV}{\ensuremath{ \d{V} }}
\newcommand{\dxis}{\ensuremath{ \d{\xi} }}
\newcommand{\lda}{\stackrel{\triangle}{\da}}

\newcommand{\dr}{\ensuremath{ \d{r} }}
\newcommand{\dphi}{\ensuremath{ \d{\phi} }}
\newcommand{\dz}{\ensuremath{\d{z}}}
\newcommand{\du}{\ensuremath{\d{u}}}
\newcommand{\dy}{\ensuremath{\d{y}}}
\newcommand{\dxscal}{\ensuremath{\d{x}}}

\renewcommand{\cos}[1]{ \text{cos}\hspace{0.0cm}\left( {#1} \right) }
\renewcommand{\sin}[1]{ \text{sin}\hspace{0.0cm}\left( {#1} \right) }
\renewcommand{\ln}[1]{\text{$\hspace{0.1cm}$ln$\left(#1\right)$}}
\renewcommand{\exp}[1]{\ensuremath{ \,\text{exp}{\left( #1 \right)} }}

\newcommand{\define}{\ensuremath{\stackrel{\mathrm{def}}{=}}}
\newcommand{\p}{\ensuremath{ ^{\prime} }}
\newcommand{\pp}{\ensuremath{ ^{\prime \prime} }}
\newcommand{\first}{$\ensuremath{ 1^{\text{st}} }\,$}
\newcommand{\second}{$\ensuremath{ 2^{\text{nd}} }\,$}
\newcommand{\lin}[1]{\ensuremath{ \calL[#1] }}

\newcommand{\tenfour}[1]{ \ensuremath {\boldsymbol{\mathcal{#1}} } }
\newcommand{\tr}[1]{\text{$\hspace{0.1cm}$tr$\left(#1\right)$}}
\newcommand{\dev}{\ensuremath{ ^{\prime} }}
\renewcommand{\det}[1]{\text{$\hspace{0.1cm}$det$\left(#1\right)$}}
\newcommand{\inv}{\ensuremath{ ^{-1} }}

\renewcommand{\it}{\ensuremath{ ^{-T} }}
\newcommand{\sym}{\ensuremath{ ^{\text{sym}} }}
\newcommand{\skw}{\ensuremath{ ^{\text{skw}} }}
\newcommand{\adj}{\ensuremath{ ^{\sharp}  }}
\newcommand{\mg}[1]{\ensuremath{ \left\| #1 \right\| }}
\newcommand{\mgv}[1]{\ensuremath{ \left| #1 \right| }}
\newcommand{\s}{\ensuremath{ ^{2}  }}
\newcommand{\ione}[1]{\ensuremath{ I_{#1} }}
\newcommand{\itwo}[1]{\ensuremath{ I\hspace{-0.1cm}I_{#1} }}
\newcommand{\ithree}[1]{\ensuremath{ I\hspace{-0.1cm}I\hspace{-0.1cm}I_{#1} }}
\newcommand{\ionep}[1]{\ensuremath{ I_{#1}\p }}
\newcommand{\itwop}[1]{\ensuremath{ I\hspace{-0.1cm}I_{#1}\p }}
\newcommand{\ithreep}[1]{\ensuremath{ I\hspace{-0.1cm}I\hspace{-0.1cm}I_{#1}\p }}
\newcommand{\ionepp}[1]{\ensuremath{ I_{#1}\pp }}
\newcommand{\itwopp}[1]{\ensuremath{ I\hspace{-0.1cm}I_{#1}\pp }}
\newcommand{\ithreepp}[1]{\ensuremath{ I\hspace{-0.1cm}I\hspace{-0.1cm}I_{#1}\pp }}

\newcommand{\gus}{\frac{\partial\scag}{\partial\tensig}}
\newcommand{\guss}{\frac{\partial^2\scag}{\partial^2\tensig}}

\newcommand{\percent}{\ensuremath{ \%  }}
\newcommand{\IE}{\ensuremath{I\hspace{-0.12cm}E  }}
\newcommand{\II}{\ensuremath{1\hspace{-0.12cm}1  }}
\newcommand{\tenIE}{\ensuremath{ \ten{\IE}  }}
\newcommand{\tenII}{\ensuremath{ \ten{\II}  }}
\newcommand{\hattenIE}{\ensuremath{ \widehat{\tenIE}  }}
\newcommand{\IC}{\ensuremath{C\hspace{-0.22cm}C  }}
\newcommand{\cc}{\ensuremath{c\hspace{-0.22cm}c  }}
\newcommand{\tenIC}{\ensuremath{ \ten{\IC}  }}
\newcommand{\tencc}{\ensuremath{ \ten{\cc}  }}
\newcommand{\lsup}[1]{\ensuremath{ {}^{[#1]} \hspace{-0.05cm} }}
\newcommand{\lsub}[1]{\ensuremath{ {}_{[#1]} \hspace{-0.05cm} }}
\newcommand{\lsupb}[2]{\ensuremath{ {}^{#1}_{#2} \hspace{-0.05cm} }}
\newcommand{\Phihat}{\ensuremath{ \widehat{\Phi} }}
\newcommand{\Psihat}{\ensuremath{ \widehat{\Psi} }}
\renewcommand{\s}{\ensuremath{ ^*} }
\renewcommand{\k}{\ensuremath{ ^K} }
\newcommand{\kp}{\ensuremath{ ^{K+1}} }
\newcommand{\km}{\ensuremath{ ^{K-1}} }
\newcommand{\tol}[1]{\ensuremath{ \text{TOL}_{#1} } }
\newcommand{\pen}{\ensuremath{ \calK }}
\newcommand{\tenone}{\ensuremath{ \ten{1} }}
\newcommand{\err}{\ensuremath{ \pounds }}
\newcommand{\li}{\ensuremath{\left[\hspace{-0.15cm}\left[ }}
\newcommand{\ri}{\ensuremath{\right]\hspace{-0.15cm}\right] }}
\renewcommand{\t}{\ensuremath{ ^{\scaT} }}
\newcommand{\tria}{\ensuremath{ ^{tr} }}

\newcommand{\iso}[1]{\ensuremath{ \widetilde{#1} }}
\newcommand{\tenFisox}{\ensuremath{ \tenF_{\vecx} } }
\newcommand{\tenFisoX}{\ensuremath{ \tenF_{\vecX} } }
\newcommand{\tenCisox}{\ensuremath{ \tenC_{\vecx} } }
\newcommand{\tenCisoX}{\ensuremath{ \tenC_{\vecX} } }
\newcommand{\Jisox}{\ensuremath{ J_{\vecx} } }
\newcommand{\JisoX}{\ensuremath{ J_{\vecX} } }
\newcommand{\calRiso}{ \ensuremath{ \iso{\calR}}}
\newcommand{\dzeta}{\ensuremath{ \d{\veczeta} } }
\newcommand{\daiso}{\ensuremath{ \d{\iso{a}} }}
\newcommand{\dviso}{\ensuremath{ \d{\iso{v}} }}
\newcommand{\vecniso}{\ensuremath{ \iso{\vecn} } }

\newcommand{\dgn}{\delta \bar{g}_{\mbox{\tiny{N}}}}
\newcommand{\lgn}{\Delta \bar{g}_{\mbox{\tiny{N}}}}
\newcommand{\ldgn}{\Delta\delta \bar{g}_{\mbox{\tiny{N}}}}

\newcommand{\dbxi}{\delta \bar{\xi}}
\newcommand{\lbxi}{\Delta \bar{\xi}}
\newcommand{\ldbxi}{\Delta\delta \bar{\xi}}

\newcommand{\intbco}{\int\limits_{\varphi\left(B^1\right)}}
\newcommand{\intbct}{\int\limits_{\varphi\left(B^2\right)}}
\newcommand{\intbcc}{\int\limits_{\varphi\left(B_c\right)}}
\newcommand{\intpbco}{\int\limits_{\varphi\left(\partial B^1\right)}}
\newcommand{\intpbct}{\int\limits_{\varphi\left(\partial B^2\right)}}
\newcommand{\intpbcon}{\int\limits_{\varphi\left(\partial B^1_n\right)}}
\newcommand{\intpbctn}{\int\limits_{\varphi\left(\partial B^2_n\right)}}
\newcommand{\intpbcc}{\int\limits_{\Gamma_c}}
\newcommand{\intbro}{\int\limits_{B^1}}
\newcommand{\intbrt}{\int\limits_{B^2}}
\newcommand{\intbrc}{\int\limits_{B_c}}

\newcommand{\intprco}{\int\limits_{\partial B^1}}
\newcommand{\intprct}{\int\limits_{\partial B^2}}
\newcommand{\intprcon}{\int\limits_{\partial B^1_n}}
\newcommand{\intprctn}{\int\limits_{\partial B^2_n}}
\newcommand{\intprcc}{\int\limits_{\partial B_c}}

\newcommand{\mnab}{n_{AB}}
\newcommand{\mnac}{n_{AC}}
\newcommand{\mnad}{n_{AD}}
\newcommand{\mnae}{n_{AE}}

\newcommand{\hthr}{\frac{h}{3}}
\newcommand{\hsix}{\frac{h}{6}}


\newcommand{\pna}{_{p\,n+1}}
\newcommand{\pno}{_{p\,n}}
\newcommand{\Ina}{_{I\,n+1}}
\newcommand{\Ino}{_{I\,n}}
\newcommand{\Jna}{_{J\,n+1}}
\newcommand{\Jno}{_{J\,n}}
\newcommand{\sal}{^{1\,\alpha}}
\newcommand{\sbl}{^{1\,\beta}}
\newcommand{\mal}{^{2\,\alpha}}
\newcommand{\xisg}{\left(\xi^1_{g\,n+1}\right)}
\newcommand{\ximg}{\left(\xi^2_{g\,n+1}\right)}
\newcommand{\xisp}{\left(\xi^1_{p\,n+1}\right)}
\newcommand{\ximp}{\left(\xi^2_{p\,n+1}\right)}
\newcommand{\xisgo}{\left(\xi^1_{g\,n}\right)}
\newcommand{\xiso}{\left(\xi^1_{p\,n}\right)}
\newcommand{\ximo}{\left(\xi^2_{p\,n}\right)}
\newcommand{\xise}{\left(\xi^1_g\left(\eta\right)\right)}
\newcommand{\xime}{\left(\xi^2_g\left(\eta\right)\right)}
\newcommand{\xipe}{\left(\xi^1_p\left(\eta\right)\right)}
\newcommand{\xisa}{\left(\xi^1_A\right)}
\newcommand{\xima}{\left(\xi^2_A\right)}
\newcommand{\xisq}{\left(\xi^1_Q\right)}
\newcommand{\ximq}{\left(\xi^2_Q\right)}

\newcommand{\etag}{\left(\eta\right)}
\newcommand{\xig}{\left(\xi_g\right)}

\newcommand{\xisf}{\left(\xi^1_1\right)}
\newcommand{\ximf}{\left(\xi^2_1\right)}
\newcommand{\xiss}{\left(\xi^1_2\right)}
\newcommand{\xims}{\left(\xi^2_2\right)}

\newcommand{\ana}{\ensuremath{ _{\alpha\,n+1} }}
\newcommand{\bna}{\ensuremath{ _{\beta\,n+1} }}

\newcommand{\sna}{\ensuremath{ _{s\,n+1} }}
\newcommand{\nna}{\ensuremath{ _{\mbox{\tiny{N}}\,n+1} }}
\newcommand{\tna}{\ensuremath{ _{\mbox{\tiny{T}}\,n+1} }}
\newcommand{\tnaa}{\ensuremath{_{\mbox{\tiny{T}}\alpha\,n+1\,}}}
\newcommand{\tnab}{\ensuremath{_{\mbox{\tiny{T}}\,n+1\,\beta}}}
\newcommand{\tnax}{\ensuremath{_{\mbox{\tiny{T}}\,n+1\,\xi}}}

\newcommand{\nA}{\ensuremath{ _{\mbox{\tiny{N}} A} }}
\newcommand{\Ng}{\ensuremath{ _{\mbox{\tiny{N}} g} }}
\newcommand{\tA}{\ensuremath{ _{\mbox{\tiny{T}} A} }}
\newcommand{\ta}{\ensuremath{ _{\mbox{\tiny{T}} \alpha} }}
\newcommand{\Tag}{\ensuremath{ _{\mbox{\tiny{T}} \alpha g} }}
\newcommand{\taA}{\ensuremath{ _{\mbox{\tiny{T}}\alpha A} }}
\newcommand{\tbA}{\ensuremath{ _{\mbox{\tiny{T}}\beta A} }}

\newcommand{\tana}{\ensuremath{ _{\mbox{\tiny{T}} A\,n+1} }}

\newcommand{\taana}{\ensuremath{ _{\mbox{\tiny{T}}\alpha A\,n+1} }}
\newcommand{\tbana}{\ensuremath{ _{\mbox{\tiny{T}}\beta A\,n+1} }}

\newcommand{\taano}{\ensuremath{ _{\mbox{\tiny{T}}\alpha A\,n} }}

\newcommand{\iana}{\ensuremath{ _{i A\,n+1} }}
\newcommand{\nana}{\ensuremath{ _{\mbox{\tiny{N}} A\,n+1} }}
\newcommand{\nano}{\ensuremath{ _{\mbox{\tiny{N}} A\,n} }}
\newcommand{\dana}{\ensuremath{ _{\mbox{\tiny{D}} A\,n+1} }}
\newcommand{\tno}{\ensuremath{ _{\mbox{\tiny{T}}\,n} }}
\newcommand{\tano}{\ensuremath{ _{\mbox{\tiny{T}}A\,n} }}
\newcommand{\na}{\ensuremath{ _{n+1} }}
\newcommand{\ina}{\ensuremath{ _{i\,n+1} }}
\newcommand{\jna}{\ensuremath{ _{j\,n+1} }}

\newcommand{\gna}{\ensuremath{ _{g\,n+1} }}
\newcommand{\qna}{\ensuremath{ _{q\,n+1} }}
\newcommand{\gno}{\ensuremath{ _{g\,n} }}

\newcommand{\Ana}{\ensuremath{ _{A\,n+1} }}
\newcommand{\Bna}{\ensuremath{ _{B\,n+1} }}
\newcommand{\Cna}{\ensuremath{ _{C\,n+1} }}
\newcommand{\Ano}{\ensuremath{ _{A\,n} }}
\newcommand{\Bno}{\ensuremath{ _{B\,n} }}
\newcommand{\Cno}{\ensuremath{ _{C\,n} }}

\newcommand{\nBna}{\ensuremath{ _{\mbox{\tiny{N}}B\,n+1} }}
\newcommand{\tBna}{\ensuremath{ _{\mbox{\tiny{T}}B\,n+1} }}

\newcommand{\xina}{\left(\vecxi_{n+1}\right)}
\newcommand{\xino}{\left(\vecxi_{n}\right)}
\newcommand{\xigna}{\left(\vecxi_{g\,n+1}\right)}
\newcommand{\xigno}{\left(\vecxi_{g\,n}\right)}

\newcommand{\agna}{\ensuremath{ _{\alpha\,g\,n+1} }}
\newcommand{\kagna}{\ensuremath{ _{,\alpha\,g\,n+1} }}
\newcommand{\agno}{\ensuremath{ _{\alpha\,g\,n} }}
\newcommand{\kagno}{\ensuremath{ _{,\alpha\,g\,n} }}

\newcommand{\ogna}{\ensuremath{ _{1\,g\,n+1} }}
\newcommand{\kogna}{\ensuremath{ _{,1\,g\,n+1} }}
\newcommand{\ogno}{\ensuremath{ _{1\,g\,n} }}
\newcommand{\kogno}{\ensuremath{ _{,1\,g\,n} }}

\newcommand{\tgna}{\ensuremath{ _{2\,g\,n+1} }}
\newcommand{\ktgna}{\ensuremath{ _{,2\,g\,n+1} }}
\newcommand{\tgno}{\ensuremath{ _{2\,g\,n} }}
\newcommand{\ktgno}{\ensuremath{ _{,2\,g\,n} }}

\newcommand{\aAna}{\ensuremath{ _{\alpha\,A\,n+1} }}
\newcommand{\bAna}{\ensuremath{ _{\beta\,A\,n+1} }}
\newcommand{\aBna}{\ensuremath{ _{\alpha\,B\,n+1} }}

\newcommand{\nnAna}{\ensuremath{ _{33\,A\,n+1} }}
\newcommand{\aaAna}{\ensuremath{ _{\alpha\alpha\,A\,n+1} }}
\newcommand{\abAna}{\ensuremath{ _{12\,A\,n+1} }}
\newcommand{\anAna}{\ensuremath{ _{\alpha3\,A\,n+1} }}

\newcommand{\iAna}{\ensuremath{ _{i\,A\,n+1} }}
\newcommand{\jAna}{\ensuremath{ _{j\,A\,n+1} }}
\newcommand{\kAna}{\ensuremath{ _{k\,A\,n+1} }}
\newcommand{\kBna}{\ensuremath{ _{k\,B\,n+1} }}

\newcommand{\dthetc}{\delta\bar{\vartheta}^1}
\newcommand{\dthets}{\delta\vartheta^2}
\newcommand{\dphic}{\delta\bar{\varphi}^1}
\newcommand{\dphis}{\delta\varphi^2}

\newcommand{\Dthetc}{\Delta\bar{\vartheta}^1}
\newcommand{\Dthets}{\Delta\vartheta^2}
\newcommand{\Dphic}{\Delta\bar{\varphi}^1}
\newcommand{\Dphis}{\Delta\varphi^2}

\newcommand{\dxi}{\delta\bar{\xi}}
\newcommand{\dxia}{\delta\bar{\xi}^{\alpha}}
\newcommand{\dxib}{\delta\bar{\xi}^{\beta}}
\newcommand{\dxic}{\delta\bar{\xi}^{\gamma}}
\newcommand{\dxid}{\delta\bar{\xi}^{\delta}}
\newcommand{\dxie}{\delta\bar{\xi}^{\epsilon}}
\newcommand{\dxif}{\delta\bar{\xi}^{\eta}}

\newcommand{\Dxi}{\Delta\bar{\xi}}
\newcommand{\Dxia}{\Delta\bar{\xi}^{\alpha}}
\newcommand{\Dxib}{\Delta\bar{\xi}^{\beta}}
\newcommand{\Dxic}{\Delta\bar{\xi}^{\gamma}}
\newcommand{\Dxid}{\Delta\bar{\xi}^{\delta}}
\newcommand{\Dxie}{\Delta\bar{\xi}^{\epsilon}}
\newcommand{\Dxif}{\Delta\bar{\xi}^{\eta}}

\newcommand{\Ddxi}{\Delta\delta\bar{\xi}}
\newcommand{\Ddxia}{\Delta\delta\bar{\xi}^{\alpha}}
\newcommand{\Ddxib}{\Delta\delta\bar{\xi}^{\beta}}
\newcommand{\Ddxic}{\Delta\delta\bar{\xi}^{\gamma}}
\newcommand{\Ddxid}{\Delta\delta\bar{\xi}^{\delta}}
\newcommand{\Ddxie}{\Delta\delta\bar{\xi}^{\epsilon}}
\newcommand{\Ddxif}{\Delta\delta\bar{\xi}^{\eta}}

\newcommand{\dvecaa}{\delta\bar{\veca}_{\alpha}^1}
\newcommand{\dvecab}{\delta\bar{\veca}_{\beta}^1}

\newcommand{\dvecaat}{\delta\bar{\veca}^{1\alpha}}
\newcommand{\dvecabt}{\delta\bar{\veca}^{1\beta}}

\newcommand{\Dvecaa}{\Delta\bar{\veca}_{\alpha}^1}
\newcommand{\Dvecab}{\Delta\bar{\veca}_{\beta}^1}
\newcommand{\Dvecac}{\Delta\bar{\veca}_{\gamma}^1}

\newcommand{\Dvecaat}{\Delta\bar{\veca}^{1\alpha}}
\newcommand{\Dvecabt}{\Delta\bar{\veca}^{1\beta}}

\newcommand{\dvecn}{\delta\bar{\vecn}^1}
\newcommand{\Dvecn}{\Delta\bar{\vecn}^1}
\newcommand{\Ddvecn}{\Delta\delta\bar{\vecn}^1}

\newcommand{\dvecua}{\delta\bar{\vecu}_{,\alpha}^2}
\newcommand{\dvecub}{\delta\bar{\vecu}_{,\beta}^2}
\newcommand{\dvecuc}{\delta\bar{\vecu}_{,\gamma}^2}
\newcommand{\dvecud}{\delta\bar{\vecu}_{,\vartheta}^2}
\newcommand{\dvecue}{\delta\bar{\vecu}_{,\theta}^2}

\newcommand{\Dvecua}{\Delta\bar{\vecu}_{,\alpha}^2}
\newcommand{\Dvecub}{\Delta\bar{\vecu}_{,\beta}^2}
\newcommand{\Dvecuc}{\Delta\bar{\vecu}_{,\gamma}^2}
\newcommand{\Dvecud}{\Delta\bar{\vecu}_{,\vartheta}^2}
\newcommand{\Dvecue}{\Delta\bar{\vecu}_{,\theta}^2}

\newcommand{\dvecuab}{\delta\bar{\vecu}_{,\alpha\beta}^2}
\newcommand{\dvecuac}{\delta\bar{\vecu}_{,\alpha\gamma}^2}
\newcommand{\dvecubc}{\delta\bar{\vecu}_{,\beta\gamma}^2}
\newcommand{\dvecuad}{\delta\bar{\vecu}_{,\alpha\vartheta}^2}
\newcommand{\dvecuae}{\delta\bar{\vecu}_{,\alpha\theta}^2}

\newcommand{\Dvecuab}{\Delta\bar{\vecu}_{,\alpha\beta}^2}
\newcommand{\Dvecuac}{\Delta\bar{\vecu}_{,\alpha\gamma}^2}
\newcommand{\Dvecubc}{\Delta\bar{\vecu}_{,\beta\gamma}^2}
\newcommand{\Dvecuad}{\Delta\bar{\vecu}_{,\alpha\vartheta}^2}
\newcommand{\Dvecuae}{\Delta\bar{\vecu}_{,\alpha\theta}^2}

\newcommand{\dvecus}{\delta\vecu^1}
\newcommand{\Dvecus}{\Delta\vecu^1}
\newcommand{\vecxa}{\bar{\vecx}_{,\alpha}^2}
\newcommand{\vecxb}{\bar{\vecx}_{,\beta}^2}
\newcommand{\vecxab}{\bar{\vecx}_{,\alpha\beta}^2}

\newcommand{\gthet}{g_{\vartheta}}
\newcommand{\dgthet}{\delta g_{\vartheta}}
\newcommand{\Dgthet}{\Delta g_{\vartheta}}

\newcommand{\thetg}{\theta_G}
\newcommand{\dthetg}{\delta\theta_G}
\newcommand{\Dthetg}{\Delta\theta_G}

\newcommand{\gphi}{g_{\varphi}}
\newcommand{\dgphi}{\delta g_{\varphi}}
\newcommand{\Dgphi}{\Delta g_{\varphi}}

\newcommand{\tta}{t_{T\alpha}}
\newcommand{\ttb}{t_{T\beta}}
\newcommand{\ttc}{t_{T\gamma}}

\newcommand{\ttx}{t_{T\xi}}

\newcommand{\ttat}{t_{T}^{\alpha}}

\newcommand{\Dtta}{\Delta t_{T\alpha}}
\newcommand{\Dttb}{\Delta t_{T\beta}}
\newcommand{\Dttc}{\Delta t_{T\gamma}}

\newcommand{\ttra}{t_{t\alpha}^{tr}}
\newcommand{\ttrb}{t_{t\beta}^{tr}}
\newcommand{\ttrc}{t_{t\gamma}^{tr}}

\newcommand{\ttrat}{t_{T}^{trial\alpha}}
\newcommand{\ttrbt}{t_{T}^{trial\beta}}
\newcommand{\ttrct}{t_{T}^{trial\gamma}}
\newcommand{\ttrdt}{t_{T}^{trial\vartheta}}

\newcommand{\Dttra}{\Delta t_{T\alpha}^{trial}}
\newcommand{\Dttrb}{\Delta t_{T\beta}^{trial}}
\newcommand{\Dttrc}{\Delta t_{T\gamma}^{trial}}

%% file: results/single-tracks.tex
\begin{table}[tb]
    \centering
    \caption{Comparison of the training times and other metrics for single-track scenarios between PINN and \pidon{}.}
    \begin{tabular}{c|c|c}
                    & PINN & \pidon{} \\
                    \hline
                    \hline
     Average training time per track  & 160 s & -- \\
     Total training time & 480 s & 870 s \\
     MAPE in temperature & 1.44 \% & 1.77 \% \\
     Average relative error in melt pool length & 0.54 \% & 2.52 \% \\
     Average relative error in melt pool width & 4.13 \% & 4.51 \% \\
     Average relative error in melt pool depth & 3.43 \% & 2.93 \% 
    \end{tabular}
    \label{tab:single-tracks}
\end{table}

\begin{figure}[p]
    \centering
    \begin{subfigure}{0.48\textwidth}
    \includegraphics[width=\textwidth]{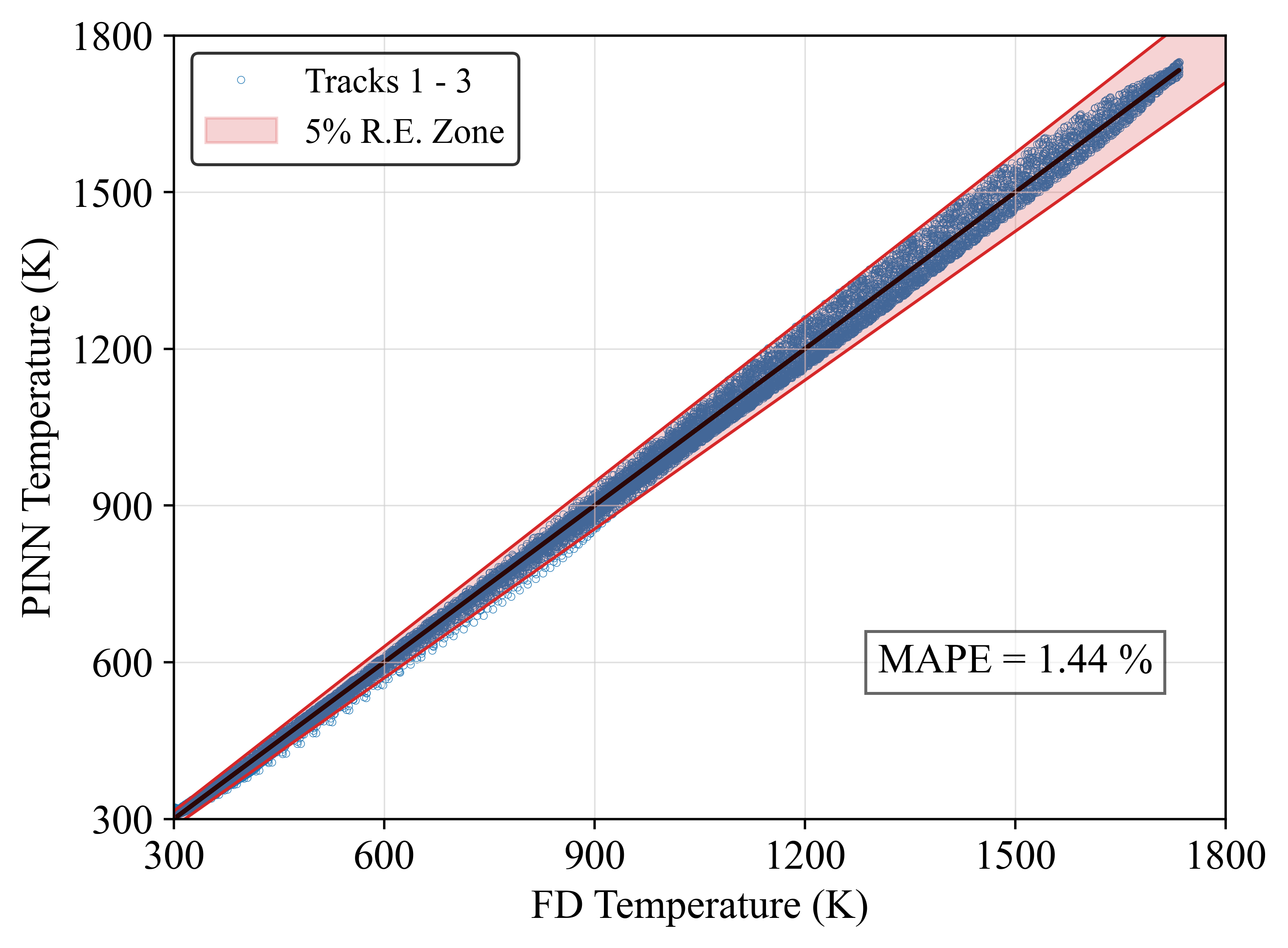}
    \caption{PINN}
    \label{fig:single-tracks_pinn}
    \end{subfigure}
    \begin{subfigure}{0.48\textwidth}
    \includegraphics[width=\textwidth]{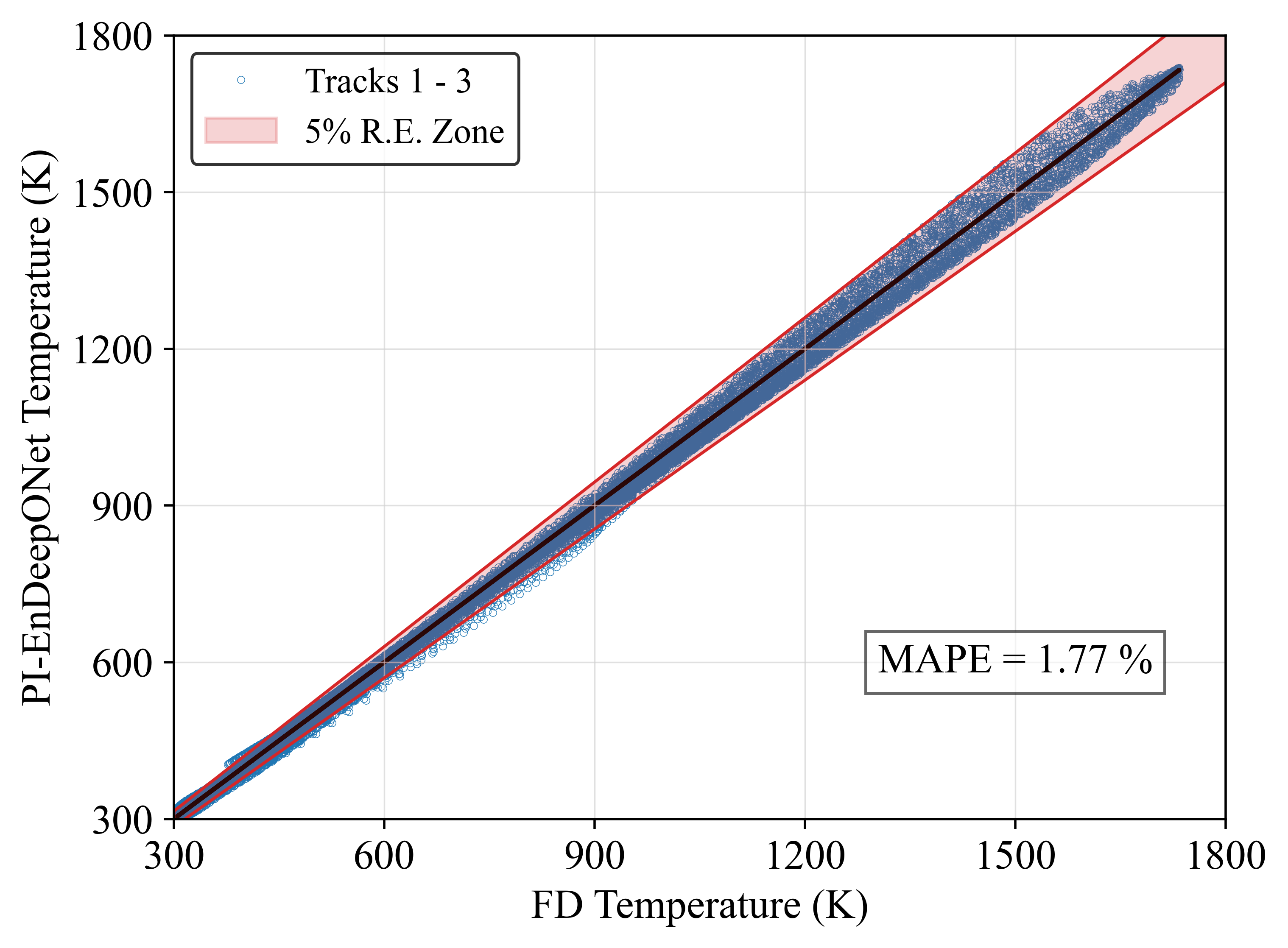}
    \caption{\pidon{}}
    \label{fig:single-tracks_pideep}
    \end{subfigure}
    \begin{subfigure}{0.98\textwidth}
    \includegraphics[width=\textwidth]{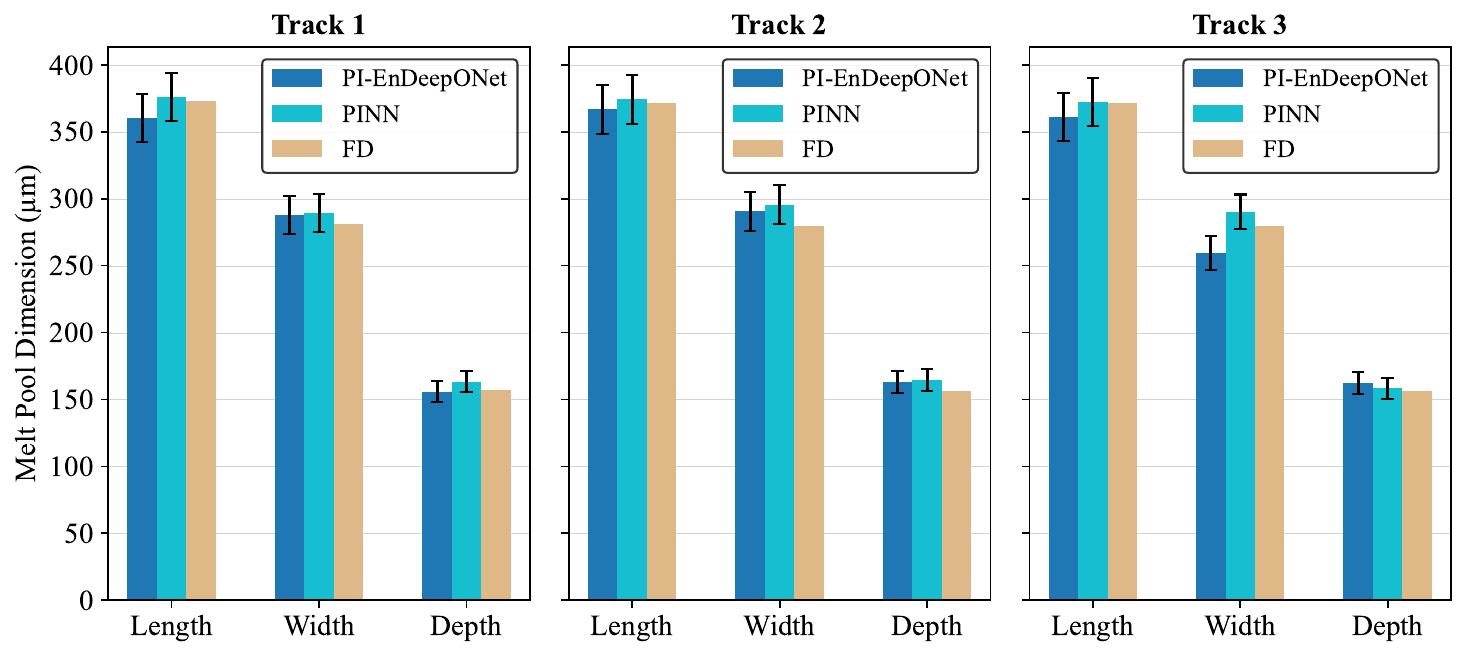}
    \caption{Melt pool dimensions}
    \label{fig:single-tracks_meltpool}
    \end{subfigure}
    \caption{Comparison of the calculated temperature and meltpool dimensions for all single-track scenarios. R.E. stands for the relative error.(a): PINN solution vs. reference (FD) solution, (b): \pidon{} solution vs. reference (FD) solution. 5 \% relative error zone is indicated by light red color. (c): Comparison of melt pool dimensions for each method.  Error bars indicate 5\% error with respect to FD solution.}
    \label{fig:single-tracks}
\end{figure}